\documentclass[1p,times]{elsarticle}
\usepackage[usenames,dvipsnames]{xcolor}
\usepackage{wrapfig}
\usepackage{amssymb}
\usepackage{amsmath}
\usepackage{amssymb}
\usepackage{amsfonts}
\usepackage{amsthm}
\usepackage{float}
\usepackage{bm}
\usepackage{latexsym}
\usepackage{hyperref}
\usepackage[capitalize]{cleveref}
\usepackage{graphicx}
\usepackage{caption}
\usepackage{subcaption}
\usepackage{tikz}
\hypersetup{
    colorlinks,
    citecolor=blue,
    linkcolor=blue,     urlcolor=blue
}

\usepackage{geometry}               	
\newcommand{\ir}{\frac{1}{r}}
\newcommand{\Integ}[1]{\int_{-\infty}^{\infty} \!\!\!\!\!\!
  \mathrm{d}#1}
\newcommand{\RInteg}[1]{\int_{0}^{\infty} \!\!\!\!\! #1\mathrm{d}#1}
\newcommand{\rInteg}{\int_{0}^{r_{max}} \!\!\!\!\! r\mathrm{d}r}
\newcommand{\TInteg}[1]{\int_{0}^{2\pi} \!\!\!\!\!\! \mathrm{d}#1}

\newcommand{\tB}[2]{\spectral{B}_{#1,m}^{#2}}
\newcommand{\tE}[2]{\spectral{E}_{#1,m}^{#2}}
\newcommand{\tj}[2]{\spectral{J}_{#1,m}^{#2}}

\newcommand{\trho}[1]{\spectral{\rho}_{m}^{#1}}

\renewcommand{\vec}[1]{\boldsymbol{#1}}

\newcommand{\spectral}[1]{\hat{\mathcal{#1}}}

\usepackage{cleveref}

\journal{Computer Physics Communications}

\begin{document}

\title{A spectral, quasi-cylindrical and dispersion-free Particle-In-Cell algorithm}
\author[label1]{R\'emi {Lehe}\corref{cor1}}
\ead{rlehe@lbl.gov}         
\author[label2]{Manuel {Kirchen}}
\author[label3]{Igor A. {Andriyash}}
\author[label1,label4]{Brendan B. {Godfrey}}
\author[label1]{Jean-Luc {Vay}}

\address[label1]{Lawrence Berkeley National Laboratory, Berkeley, CA 94720, USA}
\address[label2]{Center for Free-Electron Laser Science \& Department of Physics, University of Hamburg, 22761 Hamburg, Germany}
\address[label3]{LOA, ENSTA ParisTech, CNRS, \'Ecole polytechnique, Universit\'e Paris-Saclay, 828 bd des Mar\'echaux, 91762 Palaiseau c\'edex France}
\address[label4]{University of Maryland, College Park, MD 20742, USA}

\cortext[cor1]{Corresponding author. Tel:+1 510-486-6785}

\begin{abstract}
We propose a spectral Particle-In-Cell (PIC) algorithm that is based 
on the combination of a Hankel transform and a Fourier transform. 
For physical problems that have close-to-cylindrical symmetry, this
algorithm can be much faster than full 3D PIC algorithms. In addition,
unlike standard finite-difference PIC codes, the proposed algorithm is
free of spurious numerical dispersion, in vacuum. 
This algorithm is benchmarked in several situations that
are of interest for laser-plasma interactions. These
benchmarks show that it avoids a number of numerical artifacts, that
would otherwise affect the physics in a standard PIC algorithm -- 
including the zero-order numerical Cherenkov effect.
\end{abstract}


\maketitle

\section*{Introduction}

Particle-In-Cell (PIC) algorithms \cite{Birdsall2004,Hockney1988} 
are extensively used in several areas of physics, including the study
of astrophysical plasmas, fusion plasmas, laser-plasma
interactions and accelerator physics. Yet,
despite their wide use, PIC algorithms can be very computationally
demanding, especially in three-dimensions, and are still subject to a range of numerical artifacts. 
These shortcomings can be particularly significant when simulating 
accelerated particle beams, or 
laser-plasma interactions (such as laser-wakefield acceleration) for two reasons:
\begin{itemize}
\item 
these systems often have close-to-cylindrical
  symmetry (e.g. particle beams and laser pulses are often
  cylindrically symmetric). This prevents the use of 2D Cartesian 
or 2D cylindrical PIC algorithms (which are only well-suited for slab-like and azimuthal symmetry),
and is instead often dealt with by using
  3D Cartesian PIC algorithms, which can be very 
computationally expensive; 
\item the physical objects of interest (e.g. the laser, or the
  accelerated particle beam) often propagate close to the speed of
  light. This makes them very sensitive to \emph{spurious numerical
    dispersion}, i.e. the fact that the electromagnetic waves do not
  propagate exactly at the physical speed of light in a standard PIC
  code, but travel instead at a spuriously-altered,
  resolution-dependent velocity. In the above-mentioned cases, 
spurious numerical dispersion can lead to substantial numerical artifacts
which can mask or disrupt the physics at stake in the simulation. This
includes, for instance, numerical Cherenkov effects in general
\cite{GodfreyJCP1974}, but also more specific artifacts, such as
e.g. the erroneous prediction of the dephasing length in
laser-wakefield acceleration \cite{CowanPRSTAB2013}.

\end{itemize}
Yet several modifications can be made to the PIC algorithm, in order
to mitigate these difficulties and increase the speed and
accuracy of the simulations in these physical situations: 
\begin{itemize}
\item one of these modifications is the development of cylindrical 
PIC algorithms with azimuthal Fourier decomposition \cite{godfrey1985iprop,Lifschitz,Davidson} 
of the electromagnetic field components 
(sometimes referred to as \emph{quasi-3D} algorithms, or as
\emph{quasi-cylindrical} algorithms as we do here). By taking
into account the symmetry of the system, these algorithms can typically reduce
the cost of the simulation to a few times that of a 2D Cartesian simulation,
instead of that of a full 3D Cartesian simulation. Moreover, unlike 2D
Cartesian algorithms, these algorithms are well adapted to close-to-cylindrical
physical systems and can accurately capture physical effects that are intrinsically
3D (such as e.g. the non-linear self-focusing of an intense laser in a
plasma \cite{EsareyRMP2009});

\item a second, separate modification was introduced by the development
  of \emph{spectral} Cartesian PIC algorithms
  \cite{LinPhysFluids1974,Haber,BunemanJCP1980,DawsonRMP1983,Liu} i.e. algorithms that solve the
  Maxwell equations in Fourier space. (These algorithm are also sometimes referred to as
  \emph{pseudo-spectral algorithms} when they make use of an
  intermediate interpolation grid in real space, and this term then
  also applies to the algorithm presented here.) These algorithms contrast with
  \emph{finite-difference} algorithms, which solve the Maxwell equations
  by approximating the derivatives as finite differences on a discrete spatial grid.
Importantly, while finite-difference algorithms suffer from
spurious numerical dispersion, there is a class of spectral algorithms
-- often referred to as \emph{Pseudo-Spectral Analytical Time Domain}
(PSATD) algorithms 
\cite{Haber,BunemanJCP1980} -- which exhibits no spurious numerical
dispersion in vacuum. As a
consequence, these algorithms are free from the associated numerical
artifacts. Moreover, it was shown recently
\cite{GodfreyJCP2014,GodfreyIEEE2014,YuJCP2014,YuCPC2015} 
that spectral PIC algorithms have better stability properties when 
performing PIC simulations in
a Lorentz-boosted frame \cite{VayPRL2007,MartinsNatPhys2010,VayJCP2011}. 
These stability properties are very
promising, since boosted-frame simulations can be faster than 
their laboratory-frame counterparts by several orders of magnitude \cite{VayPRL2007}.
\end{itemize}

Even though these two improvements are both very valuable, they cannot be
combined with each other in their present
formulation. Fundamentally, this is because the
\emph{quasi-cylindrical} algorithm 
uses a cylindrical system of coordinates, whereas most spectral algorithms
(and in particular the PSATD algorithms)
were developed in a Cartesian system of coordinates. This
incompatibility is however not definitive, and the aim of this article
is to overcome these differences by developing a \emph{spectral
  quasi-cylindrical} formalism. In this document, we derive this formalism,
and we show that it can be used to build a PIC algorithm that combines 
the speed of \emph{quasi-cylindrical} algorithms with
the accuracy and lack of spurious numerical dispersion of the PSATD algorithms.

Although the algorithm described here is, to our knowledge, unique in
its capabilities, there are several other existing codes
  which have some similarities with it. One such example is the
hybrid, quasi-cylindrical version of \textsc{OSIRIS} \cite{Yuarxiv2015}. This algorithm
involves a Fourier transform in the longitudinal direction, but
retains a finite-difference formulation in the transverse (radial)
direction. As a consequence, this algorithm is not fully spectral, nor
fully dispersion-free. Another such example is the code \textsc{PlaRes} 
\cite{AndriyashJCP2015}, which is designed to simulate the physics of
free-electron lasers (FEL). This code also uses a spectral quasi-cylindrical
formalism, but it models the fields in narrow spectral intervals around the
resonant FEL frequencies, relies on the scalar and vector potentials
$\phi$ and $\vec{A}$ and uses no spatial grid. As a result,
\textsc{PlaRes} differs significantly from the standard PIC
formulation, and from the algorithm described here.

In the present article, we start by deriving the equations of our
spectral quasi-cylindrical formalism (\cref{sec:theory}). We then explain
how these equations were discretized and implemented in a fully
working PIC code (\cref{sec:implementation}), and we report on the
results of a number of benchmarks that were performed with this code 
(\cref{sec:benchmarks}). Finally, we describe two typical physical situations in
which our spectral algorithm performs better than standard 
finite-difference algorithms (\cref{sec:advantages}).

\section{Representation of the fields and continuous equations}
\label{sec:theory}

\subsection{A reminder on spectral Cartesian codes}

It is well-known that the Maxwell equations in Cartesian coordinates 
\begin{subequations}
\begin{align}
\frac{1}{c^2}\partial_t E_x = \partial_y B_z - \partial_z B_y - \mu_0  j_x \qquad&   
\partial_t B_x = -\partial_y E_z + \partial_z E_y \label{eq:CartMaxwellx} \\
\frac{1}{c^2}\partial_t E_y = \partial_z B_x - \partial_x B_z - \mu_0  j_y \qquad &   
\partial_t B_y = -\partial_z E_x + \partial_x E_z \label{eq:CartMaxwelly}  \\
\frac{1}{c^2}\partial_t E_z = \partial_x B_y - \partial_y B_x - \mu_0  j_z \qquad &   
\partial_t B_z = -\partial_x E_y + \partial_y E_x \label{eq:CartMaxwellz} 
\end{align}
\end{subequations}
can be solved by representing the fields as a sum of Fourier modes.
\begin{equation}
\label{eq:CartBwTrans}
F_u(\vec{r}) = \frac{1}{(2\pi)^{3}}\Integ{k_x} \,\Integ{k_y}\,
\Integ{k_z} \; \mathcal{F}_u(\vec{k}) \, e^{i(k_x x + k_y y + k_z z)} 
\end{equation}
with 
\begin{equation}
\label{eq:CartFwTrans}
\mathcal{F}_u(\vec{k})  = \Integ{x} \,\Integ{y}\, \Integ{z} \;
F_u(\vec{r}) \, e^{-i(k_x x + k_y y + k_z z)} 
\end{equation}
where $F$ is any of the fields $E$, $B$ or $j$, and where $u$ is
either $x$, $y$ or $z$. $\mathcal{F}$ represents the Fourier
components of $F$, which will be denoted
$\mathcal{E}$, $\mathcal{B}$ or $\mathcal{J}$ depending on whether
$F$ represents $E$, $B$ or $j$. With this representation, the
different Fourier modes decouple and the equations 
\cref{eq:CartMaxwellx,eq:CartMaxwelly,eq:CartMaxwellz} become 
\begin{subequations}
\begin{align}
\frac{1}{c^2}\partial_t \mathcal{E}_x = ik_y \mathcal{B}_z - ik_z \mathcal{B}_y - \mu_0 \mathcal{J}_x \qquad &   
\partial_t \mathcal{B}_x = -ik_y \mathcal{E}_z + ik_z \mathcal{E}_y \label{eq:CartSpectMaxwellx}\\
\frac{1}{c^2}\partial_t \mathcal{E}_y = ik_z \mathcal{B}_x - ik_x \mathcal{B}_z - \mu_0  \mathcal{J}_y \qquad &   
\partial_t \mathcal{B}_y = -ik_z \mathcal{E}_x + ik_x \mathcal{E}_z \label{eq:CartSpectMaxwelly}\\
\frac{1}{c^2}\partial_t \mathcal{E}_z = ik_x \mathcal{B}_y - ik_y \mathcal{B}_x - \mu_0 \mathcal{J}_z  \qquad &   
\partial_t \mathcal{B}_z = -ik_x \mathcal{E}_y + ik_y \mathcal{E}_x \label{eq:CartSpectMaxwellz}
\end{align}
\end{subequations}
The Fourier coefficients $\mathcal{E}$ and $\mathcal{B}$ can be then integrated in time, and
transformed back into real space using \cref{eq:CartBwTrans}. This is
the core principle of spectral Cartesian algorithms, including the PSATD algorithms.

\subsection{Spectral quasi-cylindrical representation}
\label{sec:representation}

The Fourier representation \cref{eq:CartBwTrans} is no longer the
appropriate representation when the Maxwell equations are written in cylindrical coordinates.
\begin{subequations}
\begin{align}
\frac{1}{c^2}\partial_t E_r = \ir \partial_\theta B_z - \partial_z B_\theta - \mu_0  j_r \qquad&   
\partial_t B_r = -\ir \partial_\theta E_z + \partial_z E_\theta \label{eq:CircMaxwellr} \\
\frac{1}{c^2}\partial_t E_\theta = \partial_z B_r - \partial_r B_z - \mu_0  j_\theta \qquad &   
\partial_t B_\theta = -\partial_z E_r + \partial_r E_z \label{eq:CircMaxwellt}  \\
\frac{1}{c^2}\partial_t E_z = \ir\partial_r r B_\theta - \ir\partial_\theta B_r - \mu_0  j_z \qquad & 
\partial_t B_z = -\ir\partial_r r E_\theta + \ir\partial_\theta E_r \label{eq:CircMaxwellzz} 
\end{align}
\end{subequations}
When replacing the representation \cref{eq:CartBwTrans} into the 
\cref{eq:CircMaxwellr,eq:CircMaxwellt,eq:CircMaxwellzz}, the Fourier
modes do not decouple. Instead one has to use the Fourier-Hankel
representation:
\begin{subequations}
\begin{align}
& F_z(\vec{r}) = \frac{1}{(2\pi)^2}\!\!\!\sum_{m=-\infty}^{\infty} \Integ{k_z}
\RInteg{k_\perp }\; \spectral{F}_{z,m}(k_z,k_\perp ) \; J_m(k_\perp r)\, e^{-im\theta + ik_z z} 
\label{eq:CircBwTransz} \\
& F_r(\vec{r}) = \frac{1}{(2\pi)^2}\!\!\!\sum_{m=-\infty}^{\infty} \Integ{k_z}\,\RInteg{k_\perp }\;
\left( \spectral{F}_{+,m}(k_z,k_\perp )\; J_{m+1}(k_\perp r) +\spectral{F}_{-,m}(k_z,k_\perp )\; J_{m-1}(k_\perp r)
\right)  e^{-im\theta +ik_z z}
\label{eq:CircBwTransr} \\
& F_\theta(\vec{r}) = \frac{1}{(2\pi)^2}\!\!\!\sum_{m=-\infty}^{\infty} \Integ{k_z}\,\RInteg{k_\perp }\;
i\left( \spectral{F}_{+,m}(k_z,k_\perp )\; J_{m+1}(k_\perp r) - \spectral{F}_{-,m}(k_z,k_\perp )\; J_{m-1}(k_\perp r)
\right)  e^{-im\theta +ik_z z} 
\label{eq:CircBwTranst}
\end{align}
\end{subequations}
where $F$ is either $E$, $B$ or $j$, where $J_m$ denotes the Bessel
function of order $m$, and where $\spectral{F}_{z,m}$, $\spectral{F}_{+,m}$ and
 $\spectral{F}_{-,m}$ represent the spectral components of $F$. (See
\ref{sec:CircTrans} for a derivation of the above equations.)
Conversely, the spectral components $\spectral{F}_{z,m}$, $\spectral{F}_{+,m}$ and
 $\spectral{F}_{-,m}$ are related to the real-space fields $F_r$,
 $F_\theta$ and $F_z$ by:
\begin{subequations}
\begin{align}
\spectral{F}_{z,m}(k_z,k_\perp ) &= \Integ{z} \RInteg{r}
\TInteg{\theta} \;F_z(\vec{r})\; J_m(k_\perp r) e^{im\theta
 - i k_z z} \label{eq:CircFwTransz} \\
\spectral{F}_{+,m}(k_z,k_\perp ) &= \Integ{z} \RInteg{r}
\TInteg{\theta} \;\frac{F_r (\vec{r})-iF_\theta (\vec{r})}{2}\; J_{m+1}(k_\perp r) e^{im\theta
 - i k_z z} \label{eq:CircFwTransp} \\
\spectral{F}_{-,m}(k_z,k_\perp ) &= \Integ{z} \RInteg{r}
\TInteg{\theta} \;\frac{F_r (\vec{r})+iF_\theta(\vec{r})}{2}\; J_{m-1}(k_\perp r) e^{im\theta
 - i k_z z} \label{eq:CircFwTransm} 
\end{align}
\end{subequations}
\noindent In the above equations, notice here that the Cartesian
component $F_z$ and cylindrical components
$F_r$, $F_\theta$ do not transform in the same manner, which is due
to their different behavior close to the axis. (Again, see
\ref{sec:CircTrans} for a more detailed explanation.)
Note also that scalar fields (like $\rho$) transform in the
same way as the Cartesian component $F_z$. 

When replacing \cref{eq:CircBwTransr,eq:CircBwTranst,eq:CircBwTransz} into the Maxwell equations in cylindrical
coordinates \cref{eq:CircMaxwellr,eq:CircMaxwellt,eq:CircMaxwellzz},
the different modes decouple, and the equations for the spectral
coefficients become:
\begin{subequations}
\begin{align}
\frac{1}{c^2}\partial_t \spectral{E}_{+,m} = - \frac{ik_\perp }{2}\spectral{B}_{z,m} + k_z\spectral{B}_{+,m} - \mu_0\spectral{J}_{+,m} \qquad &   
\partial_t \spectral{B}_{+,m} = \frac{ik_\perp }{2} \spectral{E}_{z,m} - k_z
\spectral{E}_{+,m} 
\label{eq:CircMaxwellp} \\
\frac{1}{c^2}\partial_t \spectral{E}_{-,m} = -\frac{ik_\perp }{2} \spectral{B}_{z,m} - k_z \spectral{B}_{-,m} - \mu_0  \spectral{J}_{-,m} \qquad &   
\partial_t \spectral{B}_{-,m} = \frac{ik_\perp }{2} \spectral{E}_{z,m} + k_z
\spectral{E}_{-,m} \label{eq:CircMaxwellm} \\
\frac{1}{c^2}\partial_t \spectral{E}_{z,m} = ik_\perp  \spectral{B}_{+,m} + ik_\perp \spectral{B}_{-,m}  - \mu_0 \spectral{J}_{z,m}  \qquad & 
\partial_t \spectral{B}_{z,m} = -ik_\perp  \spectral{E}_{+,m} - ik_\perp \spectral{E}_{-,m}  \label{eq:CircMaxwellz} 
\end{align}
\end{subequations}
(See \ref{sec:SpectMaxwell} for a derivation of these equations.)
Notice that these equations have a similar structure as the spectral
Cartesian equations \cref{eq:CartSpectMaxwellx,eq:CartSpectMaxwelly,eq:CartSpectMaxwellz}, as 
they also involve the product of the fields with the components of
$\vec{k}$, in their right-hand side. They do differ however in their
details, as evidenced by the signs, the factors 1/2 and the
presence or absence of the complex number $i$ in some terms.

Similarly, in this formalism, the conservation equations
$\vec{\nabla}\cdot\vec{E}=\rho/\epsilon_0$ and
$\vec{\nabla}\cdot\vec{B} = 0$
become
\begin{equation}
\label{eq:SpectCons}
k_\perp (\spectral{E}_{+,m} -\spectral{E}_{-,m}) + ik_z \spectral{E}_{z,m} =
\frac{\spectral{\rho}_m}{\epsilon_0} \qquad
 k_\perp (\spectral{B}_{+,m} -\spectral{B}_{-,m}) + ik_z \spectral{B}_{z,m} =
0 \end{equation}
As expected, the above conservation equations \cref{eq:SpectCons} are
preserved by the Maxwell equations
\cref{eq:CircMaxwellp,eq:CircMaxwellm,eq:CircMaxwellz}, provided
that the current satisfies $\partial_t
\rho + \vec{\nabla} \cdot \vec{j} = 0$, i.e. in spectral space:
\begin{equation}
\label{eq:SpectCharge}
\partial_t \spectral{\rho}_m + k_\perp (\spectral{J}_{+,m} -\spectral{J}_{-,m}) + ik_z
\spectral{J}_{z,m} = 0
\end{equation} 
\noindent (This can be checked, for instance, by differentiating
\cref{eq:SpectCons} in time and by using
\cref{eq:CircMaxwellp,eq:CircMaxwellm,eq:CircMaxwellz} to re-express the
time derivatives.)

As in a spectral Cartesian PIC codes, the equations
\cref{eq:CircMaxwellp,eq:CircMaxwellm,eq:CircMaxwellz}  
can be integrated in time, and the fields can then be transformed back into
real space, using \cref{eq:CircBwTransz,eq:CircBwTransr,eq:CircBwTranst}. Therefore,
the methods used to integrate the fields in time in a spectral Cartesian
code (e.g. PSATD) should be transposable to this spectral quasi-cylindrical formalism.

However, the advantage of this formalism over its Cartesian
counterpart is that the sum over $m$ in
\cref{eq:CircBwTransz,eq:CircBwTransr,eq:CircBwTranst} can generally
be truncated to only a few terms, for physical situations that have
close-to-cylindrical symmetry. (A similar truncation is done in \cite{Lifschitz}.) This is because the
different values of $m$ correspond to different azimuthal modes of the
form $e^{-im\theta}$, and because these modes are typically zero for
large values of $|m|$, in situations with close-to-cylindrical symmetry.
As a result, the representation of the fields is
reduced to a few 2D arrays $\spectral{F}_m(k_z,k_\perp )$ instead of the
Cartesian 3D arrays $\mathcal{F}(k_x,k_y,k_z)$. This reduction makes
the manipulation of the fields much more computationally efficient.

\section{Numerical implementation}
\label{sec:implementation}

\subsection{Overview of the algorithm}

The PIC algorithm described here uses the above-mentioned representation
of the fields, in order to solve the Maxwell equations and the motion
of charged particles in a finite-size simulation box.

Note that, when using spectral algorithms in a finite box, it is often
necessary to arbitrarily adopt specific boundary conditions (which may not
match the physics at stake), in order to be able to
conveniently represent the fields. For instance, in Cartesian spectral
codes, periodic boundaries are arbitrarily chosen in order to be able
to represent the fields as a discrete sum of Fourier modes. However, this is mostly
for mathematical convenience and it does not preclude the application,
at each timestep, of another type of boundary condition in real space
and \emph{within} the finite box (e.g. Perfectly Matched Layers),
before transforming the fields to spectral space (see
e.g. \cite{Vincenti2015,LeeCPC2015}). In the same spirit, here we arbitrarily
impose periodic boundary conditions along $z$, and Dirichlet
boundary conditions along $r$ (more specifically $\vec{E}(r_{max})=\vec{0}$ and
$\vec{B}(r_{max})=\vec{0}$), since in this case the fields can be
decomposed into a discrete Fourier-Bessel series
(see e.g. \cite{Guizar}). Yet again, this does not prevent the
practical application of other boundary conditions in real space. 

Although, as explained in \cref{sec:representation}, the fields are represented by a few
2D arrays, the particles are still distributed in 3D and their motion
is integrated in 3D Cartesian coordinates. As in a spectral Cartesian
code, we do not perform the current deposition and field gathering directly from the
macroparticles to the spectral space. (This is inefficient since the
deposition and gathering are local operations in real space -- i.e. affect
only the few cells next to the macroparticle -- but global operations
in spectral space -- i.e. they affect all the spectral modes
simultaneously.) Instead we use an intermediate grid where these
operations can be performed locally, and where the fields $\hat{F}_{u,m}$
are defined by
\begin{equation} 
\label{eq:IntermBwTrans}
F_u(\vec{r}) = \sum_{m=-\infty}^{\infty} \hat{F}_{u,m}(r,z)
e^{-im\theta} 
\end{equation}
\begin{equation}
\label{eq:IntermFwTrans}
\hat{F}_{u,m}(r,z) = \frac{1}{2\pi} \TInteg{\theta} \;
F_u(\vec{r})e^{im\theta}
\end{equation}
where ${F}$ is either ${E}$, ${B}$ or
${j}$ and $u$ is either $z$, $r$ or $\theta$. Notice that this representation is the
same as that of \cite{Lifschitz, Davidson}. In this
representation, the spectral decomposition in the azimuthal
direction is preserved, since the factors $e^{im\theta}$ can be efficiently computed from the
particle Cartesian positions $x,y,z$ by using the relation $e^{im\theta} = (x+iy)^m/r^m$
\cite{Lifschitz}.

After the particles deposit their charge and current onto this
intermediate grid, the fields of the grid are transformed into spectral
space where, as mentioned in \cref{sec:theory}, the Maxwell equations 
can be easily integrated. From \cref{eq:IntermFwTrans,eq:IntermBwTrans} and
\cref{eq:CircFwTransz,eq:CircFwTransp,eq:CircFwTransm,eq:CircBwTransz,eq:CircBwTransr,eq:CircBwTranst}, 
the transformation between the intermediate grid $\hat{F}_{m}(r,z)$
and the spectral grid $\spectral{F}_m(k_\perp, k_z)$ is:
\begin{subequations}
\begin{align}
\spectral{F}_{z,m}(k_\perp,k_z) & = \mathrm{HT}_{m} [ \; \mathrm{FT}
                               [ \; \hat{F}_{z,m}(r,z) \; ] \;] \\
\spectral{F}_{+,m}(k_\perp,k_z) &= \mathrm{HT}_{m+1}\left[ \; \mathrm{FT} \left[ \frac{
  \hat{F}_{r,m} -i  \hat{F}_{\theta,m} }{2}  \right] \;\right] \\
\spectral{F}_{-,m}(k_\perp,k_z) &= \mathrm{HT}_{m-1} \left[ \;\mathrm{FT} \left[ \frac{
  \hat{F}_{r,m} +i  \hat{F}_{\theta,m} }{2}  \right] \;\right] 
\end{align}
\end{subequations}
and
\begin{subequations}
\begin{align}
\hat{F}_{z,m}(r,z) &= \mathrm{IFT} [\; \mathrm{IHT}_{m} [
                         \spectral{F}_{z,m}(k_\perp,k_z) ] \; ] \\
\hat{F}_{r,m}(r,z) & = \mathrm{IFT} \left[ \; \mathrm{IHT}_{m+1}
                         [ \spectral{F}_{+,m}(k_\perp,k_z) ] + \mathrm{IHT}_{m-1} [
                         \spectral{F}_{-,m}(k_\perp,k_z) ] \; \right] \label{eq:FTHTr}\\
\hat{F}_{\theta,m}(r,z) & = i\;\mathrm{IFT} \left[ \; \mathrm{IHT}_{m+1}
                         [ \spectral{F}_{+,m}(k_\perp,k_z) ] -
                          \mathrm{IHT}_{m-1} [
                          \spectral{F}_{-,m}(k_\perp,k_z) ] \; \right]
\label{eq:FTHTt}
\end{align}
\end{subequations}
where $\mathrm{FT}$ represents a Fourier Transform along the $z$
axis and $\mathrm{HT}_{n}$ represents a Hankel Transform of order
$n$ along the transverse $r$ axis, and where $\mathrm{IFT}$ and
$\mathrm{IHT}_{n}$ represent the corresponding inverse
transformations:
\begin{equation} 
\mathrm{FT} [f]\, (k_z) \equiv \Integ{z} \, e^{-ik_zz} \; f(z) \qquad 
\mathrm{IFT} [g] \,(z)\equiv \frac{1}{2\pi}\Integ{k_z} \, e^{ik_zz} \;
g(k_z)
\end{equation}
\begin{equation}
\label{eq:def-Hankel} 
\mathrm{HT}_{n} [f]\,(k_\perp) \equiv 2\pi \RInteg{r} \,
J_{n}(k_\perp r) \; f(r) \qquad 
\mathrm{IHT}_{n} [g]\,(r) \equiv \frac{1}{2\pi}\RInteg{k_\perp} \,
J_{n}(k_\perp r)  \;  g(k_\perp) 
\end{equation}

The above equations show that the transformation from the intermediate
grid ($\hat{F}_{u,m}$) to the spectral grid ($\spectral{F}_{u,m}$) is
the combination of a Fourier transform (in $z$) and a Hankel transform
(in $r$). The Fourier transform in $z$ can be discretized through a Fast Fourier
Transform (FFT) algorithm, which requires an evenly-spaced grid
in $z$ and in $k_z$. On the other hand, there is more freedom of
choice for the Discrete Hankel Transform (DHT), and the implementation that
we chose is described in the next section
(\cref{sec:discretization}) and in the \ref{sec:HTMatrix}.

\begin{figure}


\begin{tikzpicture}
\def \Dt{2.8}
\def \yspac{0.7}

\begin{scope}[yshift=0cm]

\draw (-0.5,4.5*\yspac) node[anchor=west]{\textbf{a) Field gathering}};

\draw[thick,->,>=stealth] (1,0) -- (5*\Dt,0) node[below]{$t$};
\draw[thick,dashed] (-0.5,0) -- (1,0);
\draw (0.5,3*\yspac) node[red,text width=2.5cm]{\textbf{Spectral grid ($k_\perp$, $k_z$)}};
\draw (0.5,\yspac) node[red,text width=2.5cm]{\textbf{Intermediate grid ($r$, $z$)}};
\draw (0.5,-\yspac) node[blue]{\textbf{Macroparticles}};
\foreach \n in {1,3} 
\draw[fill=black] (\n*\Dt,0) circle(0.1);
\foreach \n in {2,4} 
\draw[fill=white] (\n*\Dt,0) circle(0.1);
\draw[->,>=stealth,thick] (2.2*\Dt,\yspac) .. controls (2.3*\Dt,0) .. (2.2*\Dt,-1.5*\yspac);
\draw[->,>=stealth,thick] (1.9*\Dt,-1*\yspac) .. controls (1.8*\Dt,-1*\yspac) .. (1.8*\Dt,-1.5*\yspac);
\draw (\Dt,3*\yspac) node[gray,fill=white,draw=white,rounded corners]{$ \spectral{J}^{n-1/2}_m$};
\draw (\Dt,1*\yspac) node[gray,fill=white,draw=white,rounded corners]{$ \hat{j}^{n-1/2}_m$};
\draw (\Dt,-\yspac) node[blue]{$\vec{p}_k^{n-1/2}$};
\draw (1.8*\Dt,3*\yspac) node[red,fill=white,draw=white,rounded corners]{$\spectral{\rho}^n_m$};
\draw (2.2*\Dt,3*\yspac) node[gray,fill=white,draw=white,rounded corners]{$  \spectral{B}^{n}_m, \spectral{E}^{n}_m$};
\draw (1.8*\Dt,\yspac) node[gray,fill=white,draw=white,rounded corners]{$\hat{\rho}^n_m$};
\draw (2.2*\Dt,\yspac) node[red,fill=white,draw=red,rounded corners]{$\hat{B}^{n}_m, \hat{E}^{n}_m$};
\draw (2*\Dt,-\yspac) node[blue,fill=white,draw=blue,rounded corners]{$\vec{x}_k^{n}$};
\draw (2*\Dt,-2*\yspac) node[blue,fill=white!75!blue,draw=blue,rounded corners]{$\vec{E}(\vec{x}_k), \vec{B}(\vec{x}_k)$};
\draw (3*\Dt,3*\yspac) node[gray,fill=white,draw=white,rounded corners]{$ \spectral{J}^{n+1/2}_m$};
\draw (3*\Dt,1*\yspac) node[gray,fill=white,draw=white,rounded corners]{$ \hat{j}^{n+1/2}_m$};
\draw (3*\Dt,-\yspac) node[gray]{$\vec{p}_k^{n+1/2}$};
\draw (3.8*\Dt,3*\yspac) node[gray,fill=white,draw=white,rounded corners]{$\spectral{\rho}^{n+1}_m$};
\draw (4.4*\Dt,3*\yspac) node[gray,fill=white,draw=white,rounded corners]{$  \spectral{B}^{n+1}_m, \spectral{E}^{n+1}_m$};
\draw (3.8*\Dt,\yspac) node[gray,fill=white,draw=white,rounded corners]{$\hat{\rho}^{n+1}_m$};
\draw (4.4*\Dt,\yspac) node[gray,fill=white,draw=white,rounded corners]{$\hat{B}^{n+1}_m, \hat{E}^{n+1}_m$};
\draw (4*\Dt,-\yspac) node[gray]{$\vec{x}_k^{n+1}$};

\end{scope}

\begin{scope}[yshift=-6cm]

\draw (-0.5, 4.5*\yspac) node[anchor=west]{\textbf{b) Equations of motion}};
\draw[thick,->,>=stealth] (1,0) -- (5*\Dt,0) node[below]{$t$};
\draw[thick,dashed] (-0.5,0) -- (1,0);
\draw (0.5,3*\yspac) node[red,text width=2.5cm]{\textbf{Spectral grid ($k_\perp$, $k_z$)}};
\draw (0.5,\yspac) node[red,text width=2.5cm]{\textbf{Intermediate grid ($r$, $z$)}};
\draw (0.5,-\yspac) node[blue]{\textbf{Macroparticles}};
\foreach \n in {1,3} 
\draw[fill=black] (\n*\Dt,0) circle(0.1);
\foreach \n in {2,4} 
\draw[fill=white] (\n*\Dt,0) circle(0.1);
\draw[->,>=stealth,thick] (1*\Dt,-\yspac) .. controls (2*\Dt,-1.9*\yspac) .. (2.75*\Dt,-1.2*\yspac);
\draw[->,>=stealth,thick] (2.1*\Dt,-1*\yspac) .. controls (3*\Dt,-0.7*\yspac) .. (3.8*\Dt,-1.*\yspac);
\draw (\Dt,3*\yspac) node[gray,fill=white,draw=white,rounded corners]{$ \spectral{J}^{n-1/2}_m$};
\draw (\Dt,1*\yspac) node[gray,fill=white,draw=white,rounded corners]{$ \hat{j}^{n-1/2}_m$};
\draw (\Dt,-\yspac) node[blue,fill=white,draw=blue,rounded corners]{$\vec{p}_k^{n-1/2}$};
\draw (1.8*\Dt,3*\yspac) node[red,fill=white,draw=white,rounded corners]{$\spectral{\rho}^n_m$};
\draw (2.2*\Dt,3*\yspac) node[gray,fill=white,draw=white,rounded corners]{$  \spectral{B}^{n}_m, \spectral{E}^{n}_m$};
\draw (1.8*\Dt,\yspac) node[gray,fill=white,draw=white,rounded corners]{$\hat{\rho}^n_m$};
\draw (2.2*\Dt,\yspac) node[red,fill=white,draw=white,rounded corners]{$\hat{B}^n_m, \hat{E}^{n}_m$};
\draw (2*\Dt,-\yspac) node[blue,fill=white,draw=blue,rounded corners]{$\vec{x}_k^{n}$};
\draw (2*\Dt,-2*\yspac) node[blue,fill=white,draw=blue,rounded corners]{$\vec{E}(\vec{x}_k), \vec{B}(\vec{x}_k)$};
\draw (3*\Dt,3*\yspac) node[gray,fill=white,draw=white,rounded corners]{$ \spectral{J}^{n+1/2}_m$};
\draw (3*\Dt,1*\yspac) node[gray,fill=white,draw=white,rounded corners]{$ \hat{j}^{n+1/2}_m$};
\draw (3*\Dt,-\yspac) node[blue,fill=white!75!blue,draw=blue,rounded corners]{$\vec{p}_k^{n+1/2}$};
\draw (3.8*\Dt,3*\yspac) node[gray,fill=white,draw=white,rounded corners]{$\spectral{\rho}^{n+1}_m$};
\draw (4.4*\Dt,3*\yspac) node[gray,fill=white,draw=white,rounded corners]{$  \spectral{B}^{n+1}_m, \spectral{E}^{n+1}_m$};
\draw (3.8*\Dt,\yspac) node[gray,fill=white,draw=white,rounded corners]{$\hat{\rho}^{n+1}_m$};
\draw (4.4*\Dt,\yspac) node[gray,fill=white,draw=white,rounded corners]{$\hat{B}^{n+1}_m, \hat{E}^{n+1}_m$};
\draw (4*\Dt,-\yspac) node[blue,fill=white!75!blue,draw=blue,rounded corners]{$\vec{x}_k^{n+1}$};

\end{scope}

\begin{scope}[yshift=-12cm]

\draw (-0.5,4.5*\yspac) node[anchor=west]{\textbf{c) Current and charge deposition}};

\draw[thick,->,>=stealth] (1,0) -- (5*\Dt,0) node[below]{$t$};
\draw[thick,dashed] (-0.5,0) -- (1,0);
\draw (0.5,3*\yspac) node[red,text width=2.5cm]{\textbf{Spectral grid ($k_\perp$, $k_z$)}};
\draw (0.5,\yspac) node[red,text width=2.5cm]{\textbf{Intermediate grid ($r$, $z$)}};
\draw (0.5,-\yspac) node[blue]{\textbf{Macroparticles}};
\foreach \n in {1,3} 
\draw[fill=black] (\n*\Dt,0) circle(0.1);
\foreach \n in {2,4} 
\draw[fill=white] (\n*\Dt,0) circle(0.1);
\draw[->,>=stealth,thick] (2*\Dt,-\yspac) -- (2.74*\Dt,0.8*\yspac);
\draw[->,>=stealth,thick] (2.9*\Dt,-\yspac) -- (2.9*\Dt,0.5*\yspac);
\draw[->,>=stealth,thick] (2.9*\Dt,1.4*\yspac) -- node[anchor=west]{FFT, DHT}(2.9*\Dt,2.5*\yspac);
\draw[->,>=stealth,thick] (4*\Dt,-\yspac) -- (3.26*\Dt,0.8*\yspac);
\draw[->,>=stealth,thick] (3.9*\Dt,-\yspac) -- (3.9*\Dt,0.5*\yspac);
\draw[->,>=stealth,thick] (3.9*\Dt,1.4*\yspac) -- node[anchor=west]{FFT, DHT}(3.9*\Dt,2.5*\yspac);
\draw (\Dt,3*\yspac) node[gray,fill=white,draw=white,rounded corners]{$ \spectral{J}^{n-1/2}_m$};
\draw (\Dt,1*\yspac) node[gray,fill=white,draw=white,rounded corners]{$ \hat{j}^{n-1/2}_m$};
\draw (\Dt,-\yspac) node[gray,fill=white,draw=white,rounded corners]{$\vec{p}_k^{n-1/2}$};
\draw[->,>=stealth,thick] (1.8*\Dt,3.2*\yspac) .. controls
(2.3*\Dt,4*\yspac) .. (2.8*\Dt,3.5*\yspac);
\draw (1.8*\Dt,3*\yspac) node[red,fill=white,draw=red,rounded
corners]{$\spectral{\rho}^n_m$};
\draw (2.2*\Dt,3*\yspac) node[gray,fill=white,draw=white,rounded corners]{$  \spectral{B}^{n}_m, \spectral{E}^{n}_m$};
\draw (1.8*\Dt,\yspac) node[gray,fill=white,draw=white,rounded corners]{$\hat{\rho}^n_m$};
\draw (2.2*\Dt,\yspac) node[red,fill=white,draw=white,rounded corners]{$\hat{B}^n_m, \hat{E}^{n}_m$};
\draw (2*\Dt,-\yspac) node[blue,fill=white,draw=blue,rounded corners]{$\vec{x}_k^{n}$};
\draw (3*\Dt,4*\yspac) node[above]{Current correction};
\draw (3*\Dt,3*\yspac) node[red,fill=white!75!red,draw=red,rounded corners]{$ \spectral{J}^{n+1/2}_m$};
\draw (3*\Dt,1*\yspac) node[red,fill=white!75!red,draw=red,rounded corners]{$ \hat{j}^{n+1/2}_m$};
\draw (3*\Dt,-\yspac) node[blue,fill=white,draw=blue,rounded corners]{$\vec{p}_k^{n+1/2}$};
\draw[->,>=stealth,thick] (3.8*\Dt,3.2*\yspac) .. controls
(3.5*\Dt,4*\yspac) .. (3.2*\Dt,3.5*\yspac);
\draw (3.8*\Dt,3*\yspac) node[red,fill=white!75!red,draw=red,rounded corners]{$\spectral{\rho}^{n+1}_m$};
\draw (4.4*\Dt,3*\yspac) node[gray,fill=white,draw=white,rounded corners]{$  \spectral{B}^{n+1}_m, \spectral{E}^{n+1}_m$};
\draw (3.8*\Dt,\yspac) node[red,fill=white!75!red,draw=red,rounded corners]{$\hat{\rho}^{n+1}_m$};
\draw (4.4*\Dt,\yspac) node[gray,fill=white,draw=white,rounded corners]{$\hat{B}^{n+1}_m, \hat{E}^{n+1}_m$};
\draw (4*\Dt,-\yspac) node[blue,fill=white,draw=blue,rounded corners]{$\vec{x}_k^{n+1}$};

\end{scope}

\begin{scope}[yshift=-18cm]

\draw (-0.5,4.5*\yspac) node[anchor=west]{\textbf{d) Maxwell equations}};

\draw[thick,->,>=stealth] (1,0) -- (5*\Dt,0) node[below]{$t$};
\draw[thick,dashed] (-0.5,0) -- (1,0);
\draw (0.5,3*\yspac) node[red,text width=2.5cm]{\textbf{Spectral grid ($k_\perp$, $k_z$)}};
\draw (0.5,\yspac) node[red,text width=2.5cm]{\textbf{Intermediate grid ($r$, $z$)}};
\draw (0.5,-\yspac) node[blue]{\textbf{Macroparticles}};
\foreach \n in {1,3} 
\draw[fill=black] (\n*\Dt,0) circle(0.1);
\foreach \n in {2,4} 
\draw[fill=white] (\n*\Dt,0) circle(0.1);
\draw[->,>=stealth,thick] (2.1*\Dt,1.4*\yspac) -- node[anchor=west]{FFT, DHT}(2.1*\Dt,2.5*\yspac);
\draw[<-,>=stealth,thick] (4.2*\Dt,1.5*\yspac) -- node[anchor=west]{IFFT, IDHT}(4.2*\Dt,2.5*\yspac);
\draw[->,>=stealth,thick] (2.4*\Dt,2.8*\yspac) .. controls (3.7*\Dt,4*\yspac) .. node[above]{PSATD} (4.05*\Dt,3.6*\yspac);
\draw (\Dt,3*\yspac) node[gray,fill=white,draw=white,rounded corners]{$ \spectral{J}^{n-1/2}_m$};
\draw (\Dt,1*\yspac) node[gray,fill=white,draw=white,rounded corners]{$ \hat{j}^{n-1/2}_m$};
\draw (\Dt,-\yspac) node[gray,fill=white,draw=white,rounded corners]{$\vec{p}_k^{n-1/2}$};
\draw (1.8*\Dt,3*\yspac) node[red,fill=white,draw=red,rounded corners]{$\spectral{\rho}^n_m$};
\draw (2.2*\Dt,3*\yspac) node[red,fill=white!75!red,draw=red,rounded corners]{$  \spectral{B}^{n}_m, \spectral{E}^{n}_m$};
\draw (1.8*\Dt,\yspac) node[gray,fill=white,draw=white,rounded corners]{$\hat{\rho}^n_m$};
\draw (2.2*\Dt,\yspac) node[red,fill=white,draw=red,rounded corners]{$\hat{B}^n_m, \hat{E}^{n}_m$};
\draw (2*\Dt,-\yspac) node[gray]{$\vec{x}_k^{n}$};
\draw (3*\Dt,3*\yspac) node[red,fill=white,draw=red,rounded corners]{$ \spectral{J}^{n+1/2}_m$};
\draw (3*\Dt,1*\yspac) node[gray,fill=white,draw=white,rounded corners]{$ \hat{j}^{n+1/2}_m$};
\draw (3*\Dt,-\yspac) node[blue]{$\vec{p}_k^{n+1/2}$};
\draw (3.8*\Dt,3*\yspac) node[red,fill=white,draw=red,rounded corners]{$\spectral{\rho}^{n+1}_m$};
\draw (4.4*\Dt,3*\yspac) node[red,fill=white!75!red,draw=red,rounded corners]{$  \spectral{B}^{n+1}_m, \spectral{E}^{n+1}_m$};
\draw (3.8*\Dt,\yspac) node[gray,fill=white,draw=white,rounded corners]{$\hat{\rho}^{n+1}_m$};
\draw (4.4*\Dt,\yspac) node[red,fill=white!75!red,draw=red,rounded corners]{$\hat{B}^{n+1}_m, \hat{E}^{n+1}_m$};
\draw (4*\Dt,-\yspac) node[blue]{$\vec{x}_k^{n+1}$};

\end{scope}

\end{tikzpicture}


\caption{\label{fig:GlobalScheme}Schematic description of the 4 steps
  of a PIC
  cycle. At any given time, the quantities that
  are known are shown in color (red and blue), 
while the quantities that are unknown or have been erased from memory 
are shown in gray. The quantities that are being calculated at a given step are
  displayed with a colored background, and arrows indicate which
  quantities are used for this calculation.}
\end{figure}
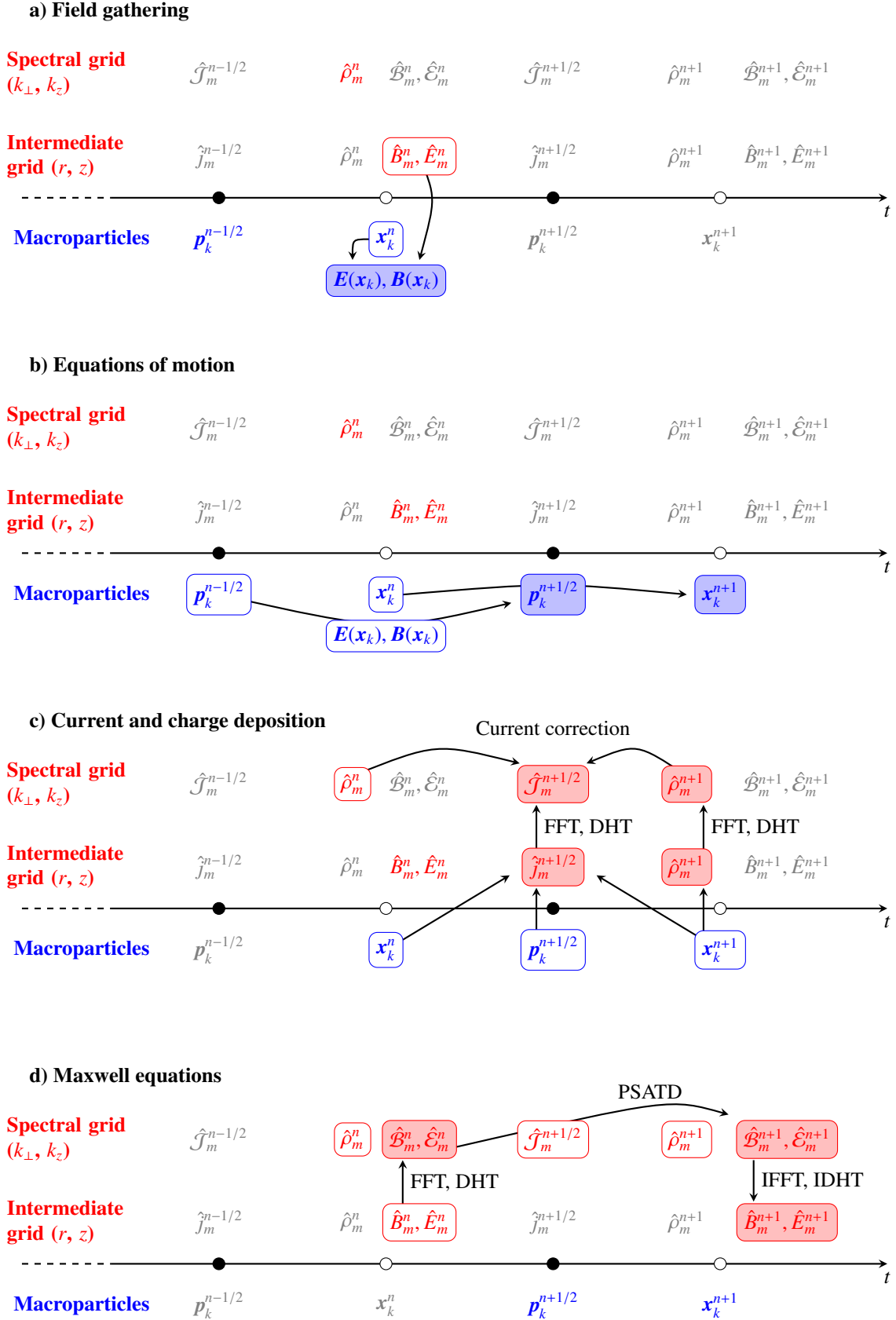

\Cref{fig:GlobalScheme} gives an overview of the successive steps
involved in one PIC cycle, including the
respective role of the intermediate grid ($\hat{F}_{u,m}$) and
spectral grid ($\spectral{F}_{u,m}$). Note that, in the PSATD
scheme that we chose (and which is described in more details in
\cref{sec:FieldIntegration}), all the fields are defined at integer
timesteps, except for the currents, which are defined at half
timesteps.  
\cref{sec:gathering,sec:eq-motion,sec:current-deposition,sec:FieldIntegration} 
describe the successive steps of the PIC cycle in more details.

\subsection{Transverse discretization of the intermediate grid, 
of the spectral grid and of the Hankel Transform}
\label{sec:discretization}

\begin{figure}[!h]
\includegraphics[width=\textwidth]{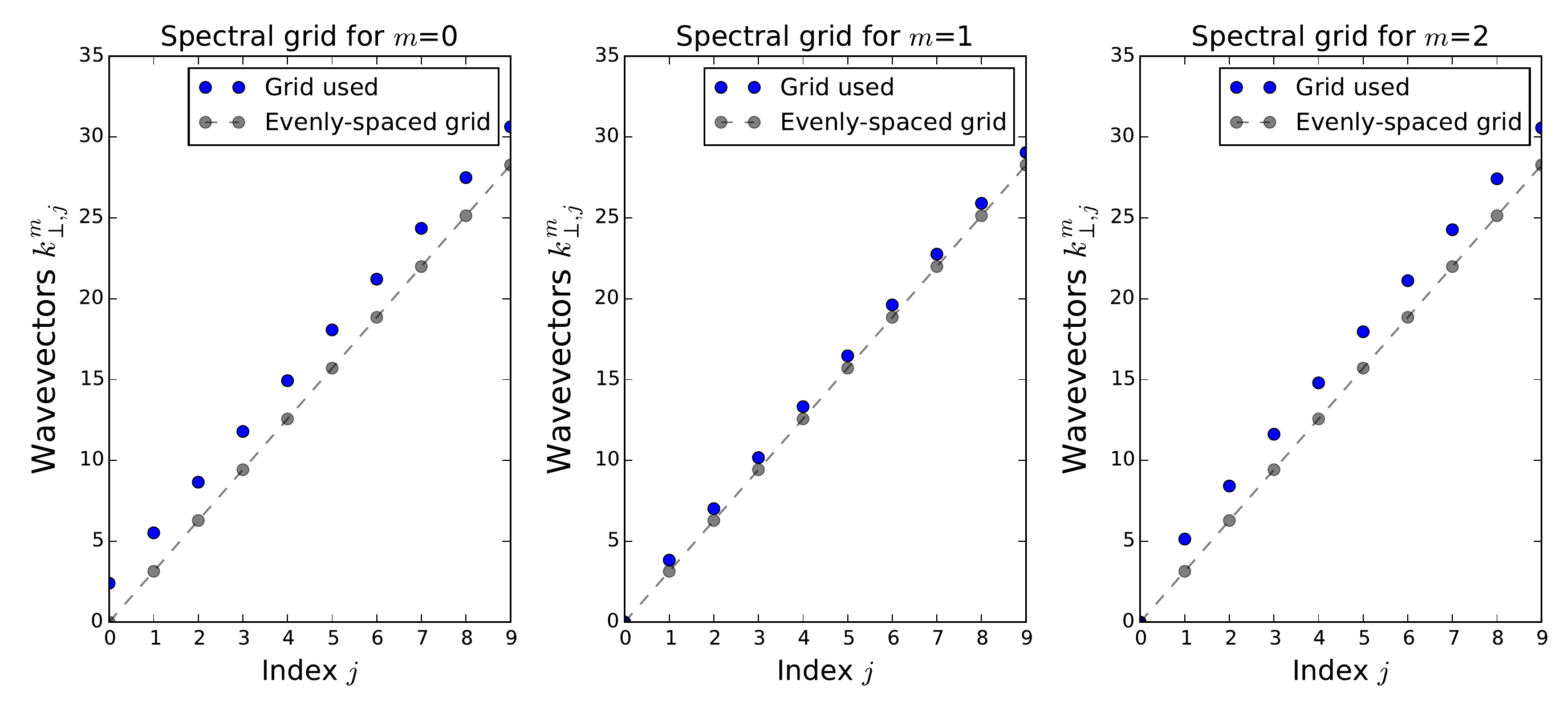}
\caption{\label{fig:Kgrid}Position of the grid points in $k_\perp$ space
  (blue dots) for the azimuthal modes $m=0$, $m=1$ and $m=2$ for $N_r
  = 10$ and $r_{max}=1$. These values are compared with those of an
  evenly-spaced grid used in spectral Cartesian codes (grey dots, $k_j = j\pi/r_{max}$).}
\end{figure}

When transforming from the intermediate to the spectral grid, the FFT
algorithm is the natural way to discretize the Fourier transform along $z$, due to its
favorable computational scaling ($\propto N_z\log(N_z)$).
On the other hand, there are a variety of existing algorithms (that are not mathematically
equivalent) to discretize the Hankel
transform (e.g. \cite{Cree,Yu,Siegman,Guizar,KaiMing}), and the use of one or the other
is very dependent on the application pursued. Broadly
speaking, choosing an algorithm for the Discrete Hankel Transform (DHT)
algorithms consists in:
\begin{itemize}
\item choosing a discrete grid in $r$ and $k_\perp$ space, on which to
  sample the functions to be transformed. In some algorithms, these
  grids may not be evenly-spaced, and can be for instance
  logarithmically spaced \cite{Siegman} or can correspond to the zeros of
  Bessel functions \cite{Yu,Guizar,KaiMing},
\item once a grid is chosen, the DHT amounts to a linear operation on a
  finite set of points (the points of the grid) and can thus be
  represented by a matrix. Thus the second choice is that of a matrix
  that represents, as closely as possible, the exact Hankel Transform.
\end{itemize}
Here, we choose to discretize the algorithm on an evenly-spaced grid in $r$
(for the intermediate grid) :
\begin{equation} 
r_j = \Delta r \left( j+\frac{1}{2} \right) \qquad  j \in \{0, ...,
N_r-1 \} \qquad \mathrm{where} \quad \Delta r = \frac{r_{max}}{N_r} 
\end{equation}
but on an irregular grid in $k_\perp$ (for the spectral grid). This is
because, for the chosen boundary condition
($\vec{E}(r_{max})=\vec{0}$, $\vec{B}(r_{max})=\vec{0}$), the fields
can be expressed as a discrete sum of Bessel modes $J_m$ with 
\begin{equation}  
k^m_{\perp,j} = \frac{\alpha_j^m}{r_{max}} \qquad j \in \{0, ...,
N_r-1 \}
\end{equation}
where $\alpha^m_j$ is the $j$th positive zero of the Bessel function of order
$m$ $J_m$ (including the trivial value $\alpha_0^m=0$ for
$m>0$; see e.g. \cite{Guizar}). 
These values are represented in figure \ref{fig:Kgrid}. Notice
that it is not an issue that the spectral components $\spectral{F}_m$
for different azimuthal modes $m$ are discretized on different
$k_\perp$ grids, since each azimuthal mode $m$ evolves separately in
\cref{eq:CircMaxwellp,eq:CircMaxwellm,eq:CircMaxwellz}.

As mentioned above, once this grid is set up, the Discrete Hankel Transform is simply a
linear operation on a finite set of points, and can thus be
represented by a matrix operation.
\begin{equation}
\mathrm{DHT^m_n}[f] \,(k^m_{\perp,j}) = \sum_{p=0}^{N_r-1} (M_{n,m})_{j,p}
\;f(r_p) \qquad \mathrm{IDHT^m_n}[g] \, (r_j) = \sum_{p=0}^{N_r-1}
(M'_{n,m})_{j,p} \; g(k^m_{\perp,p}) 
\end{equation}

Notice that the $N_r\times N_r$ transformation matrices $M_{n,m}$ and
$M_{n,m}'$ depend on $n$ (order of the Hankel transform, i.e. order of
the Bessel function $J_n$ in the integrand of \cref{eq:def-Hankel}) and $m$
(index of the azimuthal mode, and thus index of the spectral grid
$k^m_{\perp,j}$ on which the Hankel transform is performed). 
In practice, $n$ and $m$ are either equal or they differ by
$\pm 1$ (e.g. in \cref{eq:FTHTr,eq:FTHTt}, when the azimuthal mode $m$ is
transformed using the Hankel transform of order $m+1$ or $m-1$.)

The expressions of $M_{n,m}$ and $M'_{n,m}$, for the DHT algorithm
that we chose, are given in \ref{sec:HTMatrix}. 
In practice,  these matrices need to be
computed only once (at the beginning of the simulation) and can then be used at each
iteration. Notice also that the computational time for this
matrix multiplication is proportional to $N_r^2$, which is slow
compared to a 1D FFT ($\propto N_r \log(N_r)$), but
still faster than the 2D FFT ($\propto N_x N_y \log(N_x) + N_x N_y \log(N_y) $)
which is typically used in the transverse plane of a spectral 3D Cartesian code.

\subsection{Field gathering}
\label{sec:gathering}

In the following, we describe the four successive steps of a PIC cycle
with our algorithm in detail, starting with the field gathering.
When gathering the fields from the intermediate grid to the
macroparticles (see \cref{fig:GlobalScheme}),
we use the standard linear shape factors :
\begin{align} 
F_u(\vec{r}_k) &=  \sum_{p,q} S_{z,p}(z_k)S_{r,q}(r_k) \left[ \sum_{m=-N_m}^{N_m} \hat{F}_{u,m}(z_p, r_q)
  e^{-im\theta_k} \right] \nonumber \\
& = \sum_{p,q} S_{z,p}(z_k)S_{r,q}(r_k) \left[ \hat{F}_{u,0}(z_p,
  r_q) + 2\,\Re \left( \sum_{m=1}^{N_m} \hat{F}_{u,m}(z_p, r_q)
  e^{-im\theta_k} \right) \right] \label{eq:gathering}
\end{align}
where $F$ is either $E$ or $B$, $u$ is either $r$, $\theta$ or $z$, $k$ is the index of the macroparticle,
and $p$ and $q$ are the indices of the two nearest cells in $z$ and
$r$ respectively. $N_m$ is the total number
of azimuthal modes used, and $\Re$ denotes the real part of a complex
quantity. Notice that, in \cref{eq:gathering}, we used the fact that
$\hat{F}_{u,-m}(z,r) = \hat{F}^*_{u,m}(z,r) $, which can be
inferred from \cref{eq:IntermFwTrans}. Incidentally,
\cref{eq:gathering} shows that only the modes with $m\geq 0$ need to be taken into account in the code, as they are sufficient to retrieve the force on the macroparticles.

Finally, in \cref{eq:gathering}, $S_{z,p}$ and $S_{z,q}$ are the linear
shape factors in $z$ and $r$ :
\begin{subequations}
\begin{equation} 
S_{z,p}(z) = \frac{z_{p+1}- z}{\Delta z}  \qquad 
S_{z,p +1}(z) = \frac{ z - z_{p} }{\Delta z} \qquad
\mathrm{with} \quad z_{p} \leq z < z_{p +1}  
\end{equation}
\begin{equation} 
\label{eq:shape-r}
S_{r,q}(r) = \frac{ r_{q+1} - r }{  \Delta r }
\qquad S_{r,q+1}(r) = \frac{ r - r_{q} }{  \Delta r }
\qquad \mathrm{with} \quad r_{q} \leq r < r_{q+1}
\end{equation}
\end{subequations}
\noindent For particles that are in the lower half of the first
  radial cell ($r < r_0 = \Delta r/2$), \cref{eq:gathering,eq:shape-r}
  require the value $\hat{F}_{u,m}(z_p, r_{-1})$ although $r_{-1} =
  -\Delta r/2$ is not actually part of the grid. In order to still
  apply \cref{eq:gathering,eq:shape-r}, we explicitly set 
$\hat{F}_{u,m}(z_p, -\Delta r/2) = \pm \hat{F}_{u,m}(z_p,\Delta r/2)$,
where the $-$ sign is chosen whenever the considered field $\hat{F}_{u,m}$
is by definition zero on the axis, and the $+$ sign is chosen
otherwise. (e.g. $\hat{E}_{r,0}$ is by
definition zero on the axis ; see \cite{Lifschitz} for more details,
and for a similar method.)

Note that we use linear shape factors here only for
the sake of simplicity, and that higher-order shape factors
(e.g. quadratic or cubic) could also be used in principle, with
a similar mirroring of the field values across the axis.

\subsection{Equations of motion}
\label{sec:eq-motion}

Since the macroparticles evolve in 3D, we first compute
the Cartesian components $E_x$, $E_y$, $E_z$, $B_x$, $B_y$ and
$B_z$, from the fields $E_r$, $E_\theta$, $E_z$, $B_r$, $B_\theta$ and
$B_z$ that were gathered at the positions of each macroparticle. 
We then advance the equations of motion
\begin{equation} \frac{d\vec{p}}{dt} = q\vec{E} + q\vec{v}\times \vec{B} \qquad
\frac{d\vec{x}}{dt} = \frac{\vec{p}}{\gamma \,m} \end{equation}
\noindent  in standard 3D Cartesian coordinates by using the leap-frog pusher described in \cite{VayPoP2008}.

\subsection{Current deposition}
\label{sec:current-deposition}

As in \cite{Lifschitz}, the charge density is calculated on the intermediate grid in the
following way:
\begin{equation} \hat{\rho}_m(z_p,r_q) = \frac{ \sum_k  S_{z,p}(z_k)S_{r,q}(r_k) Q_k e^{im\theta_k}}{V_{q}} \end{equation}
where $Q_k$ is the charge of the macroparticle with index $k$, and
where $V_q$ is the volume of a cell, which is for our grid
\begin{equation} V_{q} = \pi [\, (q+1)^2- q^2\,] \Delta r^2 \Delta z \end{equation}

\noindent Similarly, the current deposition is given by
\begin{equation} \hat{j}_{u,m}(z_p,r_q) = \frac{\sum_k S_{z,p}(z_k) S_{r,q}(r_k)
Q_k v_{u,k} e^{im\theta_k}}{V_{q}} \end{equation}
where $u = z,r,\theta$ and the $v_{u,k}$ are the cylindrical components of the
velocity of the macroparticle $k$. As mentioned previously, once the
macroparticles have deposited their charge and current on the
intermediate grid, we transform them to the spectral grid, using an
FFT and a DHT (see \cref{fig:GlobalScheme}).

Note that, when a macroparticle travels through the grid, the factor
$V_q$ in the above formulas decreases as the macroparticle comes
closer to the axis. Therefore, close to the axis, the impact of a single macroparticle on
an individual grid cell can be substantial and this generally
translates into higher levels of noise. (This is a general
problem with cylindrical and quasi-cylindrical codes, and thus it is not specific to
the present spectral version.) In order to mitigate this problem, and
more generally in order to avoid the accumulation
of noise at high frequency, smoothing is 
typically applied on the charge and currents, after they
have been deposited (but no smoothing is applied on the
fields $E$ and $B$ directly). This smoothing
is performed directly in spectral space, by multiplying the charge
and currents by a transfer function $\spectral{T}(k_z, k_\perp)$ which
damps the high frequencies $\vec{k}$, and whose mathematical form is
identical to the spectral Cartesian representation of a single-pass
binomial filter \cite{Birdsall2004}.
\begin{equation} \spectral{T}(k_z, k_r) = \cos^2 \left( \frac{k_z}{k_{z,max}}\frac{\pi}{2} \right)
\cos^2\left( \frac{k_\perp}{k_{\perp,max}}\frac{\pi}{2} \right) \end{equation}
\noindent where $k_{z, max}$ and $k_{\perp,max}$ are the highest
wavevectors that the discrete spectral grid supports, in the
longitudinal and transverse direction.

Notice also that, in the above deposition scheme, we do not attempt to
reproduce the Esirkepov charge-conserving current deposition
\cite{Esirkepov}, and instead use a
simple direct current deposition. It is well-known that this simple deposition does not necessarily
satisfy the relation $\partial_t\rho + \vec{\nabla}\cdot\vec{j} =
0$, but that this can be corrected, by slightly modifying the
currents without modifying their curl (e.g. \cite{VayJCP2013}):
\begin{equation} \vec{j}' = \vec{j} - \vec{\nabla} G \end{equation}
where $G$ satisfies the Poisson-like equation
\begin{equation} \vec{\nabla}^2 G = \partial_t\rho + \vec{\nabla}\cdot\vec{j} \end{equation}

The above equation is typically expensive to solve on a spatial grid, but
very easy to solve in spectral space. In spectral space and with the
notations of \cref{fig:GlobalScheme}, these equations become
\begin{equation} \spectral{J}'^{\,n+1/2}_{+,m} = \spectral{J}^{n+1/2}_{+,m} +
\frac{k_\perp}{2} \spectral{G}^{n+1/2}_m
\qquad
\spectral{J}'^{\,n+1/2}_{-,m} = \spectral{J}^{n+1/2}_{-,m} - \frac{k_\perp}{2} \spectral{G}^{n+1/2}_m
\qquad \spectral{J}'^{\,n+1/2}_{z,m} = \spectral{J}^{n+1/2}_{z,m} - ik_z
\spectral{G}^{n+1/2}_m\end{equation}
with
\begin{equation} \spectral{G}^{n+1/2}_m = - \frac{1}{k_\perp^2 + k_z^2}\left(
  \frac{\spectral{\rho}^{n+1}_m -\spectral{\rho}^{n}_m}{\Delta t} + k_\perp
  (\spectral{J}^{n+1/2}_{+,m} -\spectral{J}^{n+1/2}_{-,m}) + ik_z\spectral{J}^{n+1/2}_{z,m}  \right) \end{equation}
With this correction, the new currents $\spectral{J}'^{n+1/2}$ do satisfy
the charge conservation equation \cref{eq:SpectCharge}. Therefore we apply this
correction in spectral space, at the end of the current deposition at each timestep.

\subsection{Integration of the Maxwell equation using the PSATD scheme}
\label{sec:FieldIntegration}

The Maxwell equations
\cref{eq:CircMaxwellp,eq:CircMaxwellm,eq:CircMaxwellz} could in
principle be integrated by using a finite-difference scheme in time, 
which would be an adaptation of the Cartesian PSTD scheme \cite{Liu}.
\begin{subequations}
\label{eq:PSTD1}
\begin{align}
\tB{+}{n+1/2} = \; & \tB{+}{n} - 
\frac{\Delta t}{2}\left(-\frac{ik_\perp }{2} \tE{z}{n} + k_z\tE{+}{n}
\right) & \\
\tB{-}{n+1/2} =\; & \tB{-}{n} - 
\frac{\Delta t}{2}\left(- \frac{ik_\perp }{2} \tE{z}{n} - k_z\tE{-}{n}
\right) &\\
\tB{z}{n+1/2} =\; & \tB{z}{n} - 
\frac{\Delta t}{2}\left(ik_\perp \tE{+}{n} + ik_\perp \tE{-}{n}
\right) &
\end{align}
\end{subequations}
\begin{subequations}
\label{eq:PSTD2}
\begin{align}
\tE{+}{n+1} = \; & \tE{+}{n} + 
c^2\Delta t\left(-\frac{ik_\perp }{2} \tB{z}{n+1/2} + k_z\tB{+}{n+1/2}
- \mu_0 \tj{+}{n+1/2} \right) & \\
\tE{-}{n+1} =\; & \tE{-}{n} +
c^2\Delta t\left(- \frac{ik_\perp }{2} \tB{z}{n+1/2} - k_z\tB{-}{n+1/2}
- \mu_0 \tj{-}{n+1/2} \right) &\\
\tE{z}{n+1} =\; & \tE{z}{n} + 
c^2\Delta t\left(ik_\perp \tB{+}{n+1/2} + ik_\perp \tB{-}{n+1/2}
- \mu_0 \tj{z}{n+1/2} \right)  &
\end{align}
\end{subequations}
\begin{subequations}
\label{eq:PSTD3}
\begin{align}
\tB{+}{n+1} = \; & \tB{+}{n+1/2} - 
\frac{\Delta t}{2}\left(-\frac{ik_\perp }{2} \tE{z}{n+1} + k_z\tE{+}{n+1}
\right) & \\
\tB{-}{n+1} =\; & \tB{-}{n+1/2} - 
\frac{\Delta t}{2}\left(- \frac{ik_\perp }{2} \tE{z}{n+1} - k_z\tE{-}{n+1}
\right) &\\
\tB{z}{n+1} =\; & \tB{z}{n+1/2} - 
\frac{\Delta t}{2}\left(ik_\perp \tE{+}{n+1} + ik_\perp \tE{-}{n+1}
\right) &
\end{align}
\end{subequations}
However, this type of scheme retains some amount of spurious numerical
dispersion, and can thus affect the simulated physics.

Instead, here we use an adaptation of the Cartesian PSATD scheme \cite{Haber}
for our spectral quasi-cylindrical representation. As in the case of the standard
PSATD, we assume that the currents are constant over one timestep and
that the charge density is linear in time over the same timestep. Under these
assumptions, the Maxwell equations \cref{eq:CircMaxwellp,eq:CircMaxwellm,eq:CircMaxwellz} can be integrated
analytically over that timestep, and they lead to:
\begin{subequations}
\label{eq:PSATD1}
\begin{align}
\tE{+}{n+1} = \; & C \tE{+}{n} + 
c^2\frac{S}{\omega}\left(-\frac{ik_\perp }{2} \tB{z}{n} + k_z\tB{+}{n}
- \mu_0 \tj{+}{n+1/2} \right) + \frac{c^2}{\epsilon_0}
\frac{k_\perp}{2}\left[ \frac{\trho{n+1}}{\omega^2}\left(
  1 - \frac{S}{\omega\Delta t}\right) -
\frac{\trho{n}}{\omega^2}\left( C -\frac{S}{\omega\Delta t}\right)\right]  & \\
\tE{-}{n+1} =\; & C \tE{-}{n} +
c^2\frac{S}{\omega}\left(- \frac{ik_\perp }{2} \tB{z}{n} - k_z\tB{-}{n}
- \mu_0 \tj{-}{n+1/2} \right) - \frac{c^2}{\epsilon_0}
\frac{k_\perp}{2}\left[ \frac{\trho{n+1}}{\omega^2}\left(
  1 - \frac{S}{\omega\Delta t}\right) - \frac{\trho{n}}{\omega^2}
\left( C - \frac{S}{\omega\Delta t}\right)\right]  &\\
\tE{z}{n+1} =\; & C \tE{z}{n} + 
c^2\frac{S}{\omega}\left(ik_\perp \tB{+}{n} + ik_\perp \tB{-}{n}
- \mu_0 \tj{z}{n+1/2} \right) - \frac{c^2}{\epsilon_0}
ik_z\left[ \frac{\trho{n+1}}{\omega^2}\left(
  1 - \frac{S}{\omega\Delta t}\right) - \frac{\trho{n}}{\omega^2}
\left( C - \frac{S}{\omega\Delta t}\right)\right]  &
\end{align}
\end{subequations}
\begin{subequations}
\label{eq:PSATD2}
\begin{align}
\tB{+}{n+1} = \; & C \tB{+}{n} - 
\frac{S}{\omega}\left(-\frac{ik_\perp }{2} \tE{z}{n} + k_z\tE{+}{n}
\right) + \mu_0 c^2\frac{1-C}{\omega^2} \left( -\frac{ik_\perp }{2}
  \tj{z}{n+1/2} + k_z \tj{+}{n+1/2} \right)& \\
\tB{-}{n+1} =\; & C \tB{-}{n} - 
\frac{S}{\omega}\left(- \frac{ik_\perp }{2} \tE{z}{n} - k_z\tE{-}{n}
\right) + \mu_0 c^2\frac{1-C}{\omega^2} \left( - \frac{ik_\perp }{2}
  \tj{z}{n+1/2} - k_z \tj{-}{n+1/2} \right) &\\
\tB{z}{n+1} =\; & C \tB{z}{n} - 
\frac{S}{\omega}\left(ik_\perp \tE{+}{n} + ik_\perp \tE{-}{n}
\right) + \mu_0 c^2\frac{1-C}{\omega^2} \left( ik_\perp
  \tj{+}{n+1/2} + ik_\perp \tj{-}{n+1/2} \right)&
\end{align}
\end{subequations}

\noindent where $\omega \equiv c\sqrt{k_z^2 + k_\perp^2}$, $C \equiv \cos(\omega \Delta t)$
and $S \equiv \sin(\omega \Delta t) $. (See \ref{sec:PSATDderiv} for a
derivation of these equations.)

\subsection{Practical implementation}

The full PIC algorithm described in this section was implemented in the
code \textsc{FBPIC} (Fourier-Bessel Particle-In-Cell), which is
written in Python. For performance, this implementation makes use of the
pre-compiled libraries FFTW \cite{FFTW} and BLAS \cite{BLAS} (for the
matrix multiplication in the DHT), and utilizes the Numba just-in-time 
compiler \cite{Numba} for the computationally-intensive parts of the
code (current deposition and field gathering). In addition, the code
was developed for both single-CPU and single-GPU architectures 
(with the GPU runs being typically more than 40 times faster than the 
equivalent CPU runs, on modern hardware). Importantly, a precise
timing of the different routines showed that, although the time taken
by the spectral transforms (FFT and DHT) is not entirely negligible,
it does not usually dominate the PIC cycle. For instance, on a K20
GPU and for a grid with $N_z=4096$, $N_r=256$ and 16 particles per cell, the FFTs 
and DHTs take up 6\% and 14\% of the PIC cycle respectively, while the
rest of the time is dominated by the field gathering and current
deposition.

Notice that, even though parallelization is generally challenging for spectral
algorithms, a multi-CPU/multi-GPU version could still be developped in the
future by using the method of \cite{VayJCP2013}. For the present
algorithm, this would involve a domain decomposition along the $z$
axis, whereby FFTs would be performed locally within each subdomain 
and whereby a large number of guard cells would be used in order to
mitigate the errors at the border between subdomains. This method
based on local FFTs has been studied in the Cartesian
context \cite{Vincenti2015}, and is now rather well understood. By contrast, 
performing domain decomposition in the $r$ direction would be 
more challenging, due to the absence of
previous work on local Hankel transforms. At any rate, the above
considerations for parallel implementation are out of the scope of the present
article, and will be the subject of future work. The benchmarks
described in the following section use a single-CPU/single-GPU version
of the algorithm.

\section{Benchmarks}
\label{sec:benchmarks}

We tested the above algorithm in a number of physical situations. These
tests included standard problems, such as e.g. a periodic plasma wave,
and involved comparing simulated results to analytical solution, as well
as verifying that the algorithm conserves the energy to a satisfying
level. For the sake of conciseness, in the present article, we
restrict the discussion to the tests of the dispersion
relation, since the absence of spurious numerical dispersion was one 
of the original goals of this algorithm.

\subsection{Propagation in vacuum}
\label{sec:vacuum_vg}

A first test consisted in letting a laser pulse propagate in vacuum and in
measuring its group velocity in the simulation. These simulations were
run with the PSATD quasi-cylindrical algorithm presented in
\cref{sec:implementation} (see
esp. \cref{eq:PSATD1,eq:PSATD2}), 
but also, for comparison, with a PSTD
version of this same algorithm (see \cref{eq:PSTD1,eq:PSTD2,eq:PSTD3}), 
as well as with a finite-difference quasi-cylindrical
algorithm equivalent to that of \cite{Lifschitz,Davidson} and which had
been previously implemented in the PIC code \textsc{Warp} \cite{Warpref}.

The simulations were run with a moving window, in a box with a
longitudinal size of 40 $\mu$m and a transverse size of 48 $\mu$m. (In
the case of the spectral algorithm, the fields were damped at the back
of the moving window, in a similar way as in \cite{YuIPAC2015}, in
order to prevent periodic wrapping of the fields).
The laser pulse itself was initialized at focus, with a waist 
$w_0 = 16 \;\mathrm{\mu m}$, a length $L = 10 \; \mathrm{\mu m}$,
a wavelength $\lambda = 0.8 \; \mathrm{\mu m}$ and a 
dimensionless potential vector $a_0 = 10^{-2}$.
The resolution was varied while keeping the same cell aspect ratio ($\Delta r = 5\Delta z$). The
timestep was set to $c\Delta t = \Delta z$ in the case of the PSATD
algorithm, to $c\Delta t = 0.9\times 2/\pi\sqrt{1/\Delta z^2 +
  1/\Delta r^2}$ in the case of the PSTD algorithm, and to $c\Delta t = 1/\sqrt{1/\Delta z^2 +
  2/\Delta r^2}$ in the case of the finite-difference algorithm (for
the PSTD and finite-difference algorithms, the chosen $\Delta t$ is
close to the Courant limit). 

Physically, the on-axis group velocity of the pulse should be slightly lower than $c$ due
to the finite waist of the pulse (see e.g. \cite{Esarey1999}). The
analytical expression for the corresponding relative
difference in on-axis group velocity is
\begin{equation} 
\label{eq:vacuum_vg}
\frac{c-v_g}{c} = 2\left( \frac{\lambda}{2\pi w_0} \right)^2
\end{equation}
\noindent In the case at hand ($w_0=16\;\mathrm{\mu m}$,
$\lambda=0.8\;\mathrm{\mu m}$), this relative difference is extremely
small ($(c-v_g)/c = 1.27 \times 10^{-4}$), and it may be difficult for
PIC codes to capture this small physical difference. 

\begin{figure}[!h]
\centering
\includegraphics[width=0.6\textwidth]{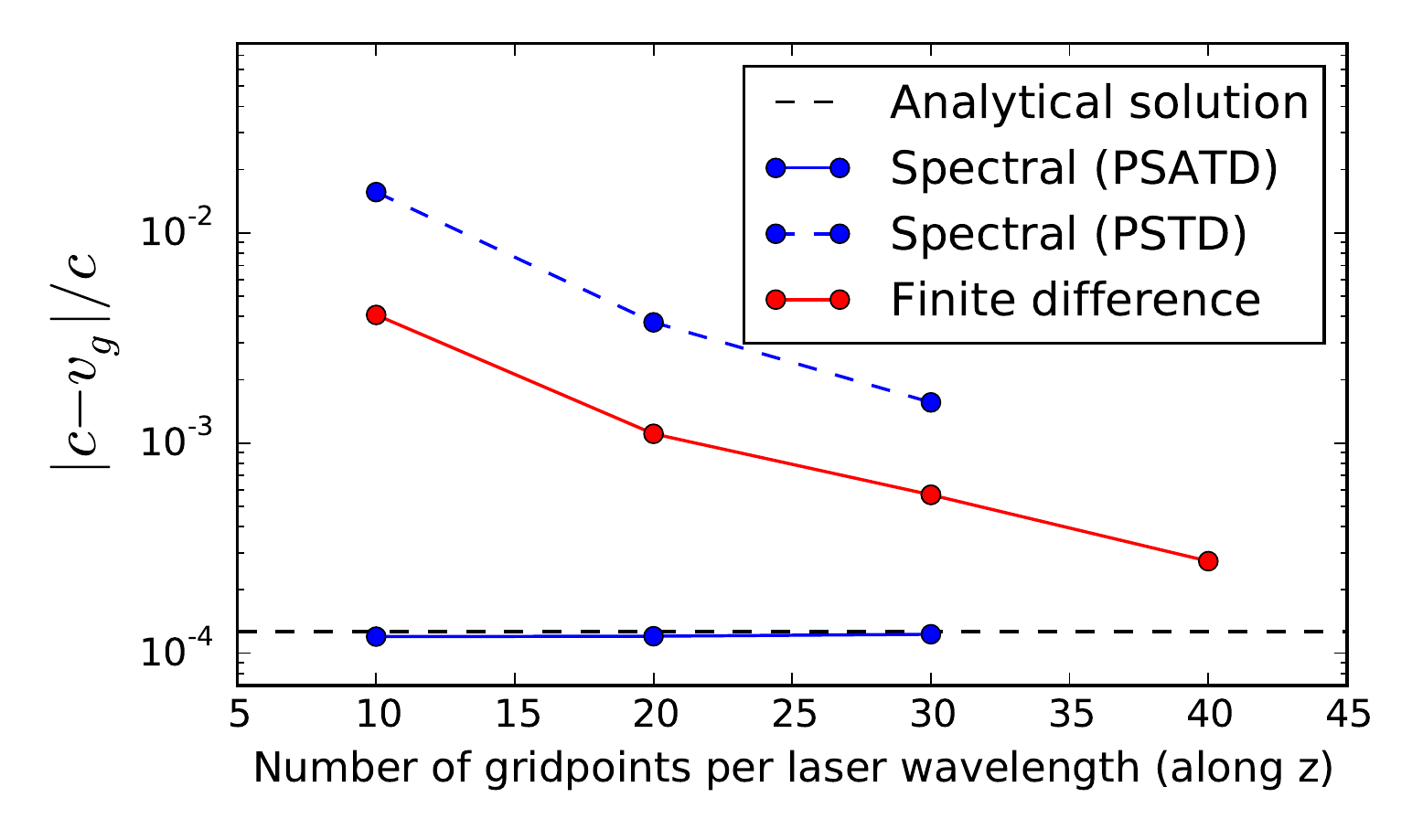}
\caption{\label{fig:Vacuum_vg}Relative difference between $c$ and the
group velocity of a laser pulse in vacuum, for different
resolutions. The dashed black line represents
the analytical prediction given by \cref{eq:vacuum_vg}, while the
blue and red points represent the results of PIC simulation, with the
PSATD, PSTD and finite-difference algorithm.}
\end{figure}

To assess the capacities of the finite-difference and
spectral codes in this regard, \cref{fig:Vacuum_vg} displays the relative
difference in group velocity, as measured in the
simulations. As can be observed, in the
finite-difference simulations and PSTD simulations, the
group velocity of the laser depends on the resolution, due to spurious
numerical dispersion. Moreover, in these cases, the group velocity is
considerably different than the analytical prediction, even for a
relatively high resolution. (Since \cref{fig:Vacuum_vg} does not display the sign
of $c-v_g$, it is worth noting here that the PSTD algorithm leads to
$v_g > c$ while the finite-difference algorithm has $v_g < c$. It is
also important to realize here that, although the finite-difference
algorithm performs better than the PSTD algorithm in this particular
example, this is not generalizable to other cases, as e.g. the numerical
dispersion of both algorithms behave differently when $\Delta t$ or the ratio $\Delta
z/\Delta r$ is changed.)

On the other hand, with the PSATD algorithm, the group velocity is 
practically independent of the resolution, and
displays very good agreement with the analytical prediction. 
This corroborates the fact that the PSATD quasi-cylindrical algorithm
described here has no spurious numerical dispersion in vacuum.

Since the PSTD algorithm is less acurate than the PSATD algorithm,
while not providing any substantial advantage in terms of speed or
practical implementation, we do not consider it further here and
perform the rest of the tests with the PSATD algorithm.

\subsection{Linear propagation in a plasma}
\label{sec:linear_plasma}

In order to confirm that the spectral algorithm also performs well in
the presence of a plasma, we ran the same type of simulations with a
uniform, pre-ionized plasma. The numerical parameters of the
simulations, as well as the physical parameters of the laser, 
were the same as in the previous subsection
(\cref{sec:vacuum_vg}). The plasma was represented by 16
macroparticles per cell, and had a density $n_e = 1.75\times
10^{18}\;\mathrm{cm}^{-3} = 10^{-3}\,n_c$, where $n_c$ is the critical
density for $\lambda=0.8\;\mathrm{\mu m}$.

Since the intensity of the laser is very low here ($a_0 = 10^{-2}$),
the propagation is linear, and the on-axis group velocity is given by
(e.g. \cite{Esarey1999})
\begin{equation} 
\label{eq:plasma_vg}
\frac{c-v_g}{c} = \frac{n_e}{2n_c} + 2\left( \frac{\lambda}{2\pi w_0} \right)^2
\end{equation}

\begin{figure}[!h]
\centering
\includegraphics[width=0.6\textwidth]{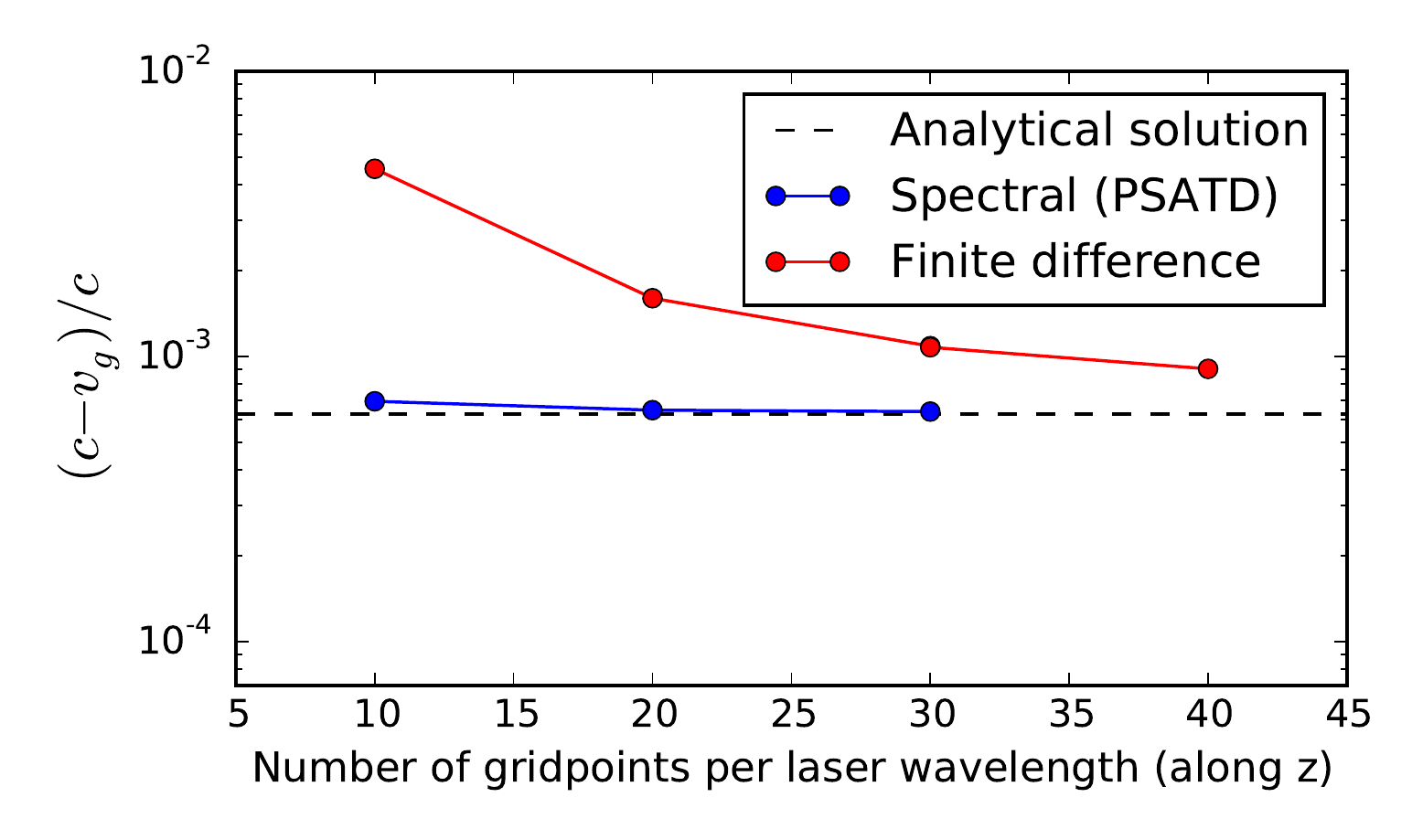}
\caption{\label{fig:Plasma_vg}Relative difference between $c$ and the
group velocity of a laser pulse in a plasma at $10^{-3}\,n_c$, for different
resolutions. The dashed line represents
the analytical prediction given by \cref{eq:plasma_vg}, while the blue
and red points represent the results of PIC simulations.}
\end{figure}

In \cref{fig:Plasma_vg}, we compare this analytical prediction with the group velocity in the
finite-difference and spectral simulations. Again, the
group velocity is resolution-dependent in the finite-difference
algorithm, and it is substantially different than the analytical
prediction even at high resolution. In the spectral code, the group
velocity velocity exhibits a very weak dependence on resolution
(which is likely due to the errors of current deposition and field
gathering on the finite grid). However, its value remains always very
close to the analytical prediction at all resolutions.

We emphasize that, although the difference between $c$ and $v_g$ is
small here and may thus seem unimportant, it is this difference which determines
the dephasing length and thus the
maximum beam energy, in a laser-wakefield simulation. It is therefore
paramount to obtain its correct value in the simulation codes. This
problem is well-known in the case of standard finite-difference
codes. For Cartesian finite-difference codes, a common solution is to
use a scheme which is dispersion-free along the $z$ axis 
(e.g. \cite{Karkkainen,Pukhov,Nuter}), but no such scheme has been
developed for quasi-cylindrical codes. Alternatively, spurious
numerical dispersion is often 
dealt with in finite-difference codes by either using an even finer grid in
$z$ (which is very computationally expensive) or
a \emph{coarser} grid in $r$. (Recall that, in the simulations shown
here, $\Delta r = 5\Delta z$. A coarser resolution in $r$ allows to
use a slightly larger timestep $\Delta t$ that approaches $c\Delta z$, the limit 
at which spurious numerical dispersion vanishes in the $z$-direction.) 
However, a coarser resolution in $r$ may not always be
adapted to resolve the physics of interest, while a finer resolution in $z$ 
is expensive. It is therefore remarkable 
that, with the spectral quasi-cylindrical algorithm, the correct group
velocity is obtained independently of the cell aspect ratio, without having to 
specifically adapt the resolution in $r$ or $z$.

\subsection{Linear laser-wakefield}

In order to further ascertain that the spectral algorithm described here gives
appropriate results beyond simple dispersion tests, we ran a simulation of a laser-wakefield, and
compared the amplitude of the wakefield with the corresponding analytical predictions.

In these simulations, the laser was linearly polarized along the
transverse $x$ direction, and had an amplitude $a_0 = 10^{-2}$, a
wavelength $\lambda=0.8 \;\mathrm{\mu m}$, a length $L=10\;\mathrm{\mu
m}$ and a waist $w_0 = 20\;\mathrm{\mu m}$. The plasma was preionized,
with a uniform electron density $n_e = 1.75\times
10^{18}\;\mathrm{cm}^{-3} = 10^{-3}\,n_c$. As is often the case for
simulations of intense laser-plasma interaction (where thermal effects
are generally assumed to be negligible), the initial temperature of
the plasma is set to 0. The simulation was run in
a moving window whose longitudinal and transverse sizes were $80 \;
\mathrm{\mu m}$ and $60 \; \mathrm{\mu m}$. The spatial and temporal
resolutions were $\Delta z = 0.05 \; \mathrm{\mu m}$, $\Delta r = 0.5
\;\mathrm{\mu m}$ and $\Delta t = \Delta z/c$. Since here $a_0 =
10^{-2}$, the analytical wakefield is given by
the linear and quasistatic theory. For a laser of the
form $a(z, r, t)= a_\ell(z, t) e^{-r^2/w_0^2} \cos(k_0z-\omega_0 t)$ --
where $a_\ell(z, t)$ is the longitudinal envelope of the laser -- the
longitudinal and transverse fields $E_z$ and $E_y$ are given by
(e.g. \cite{EsareyRMP2009})
\begin{subequations}
\begin{equation} 
E_z(z, r, t) = \frac{mc^2}{e} \frac{k_p^2}{4}\int_{z}^{\infty} 
a_\ell^2(z', t) e^{-r^2/w_0^2} \cos[k_p(z-z')]dz' \label{eq:analytical-Ez}
\end{equation}
\begin{equation}
E_y(z, r, t) = -\frac{mc^2}{e} \frac{k_p y}{w_0^2}\int_{z}^{\infty} 
a_\ell^2(z', t) e^{-r^2/w_0^2} \sin[k_p(z-z')]dz' \label{eq:analytical-Ey}
\end{equation}
\end{subequations}
\noindent where $k_p$ is the plasma wavevector.

\begin{figure}[!h]
\centering
\includegraphics[width=\textwidth]{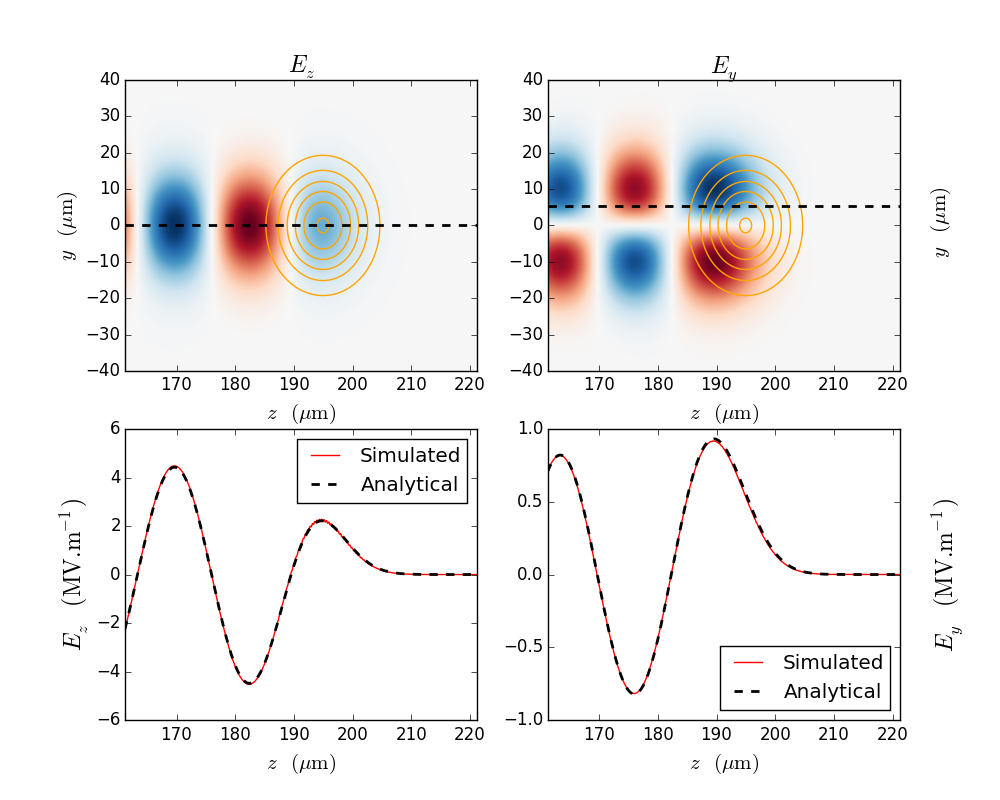}
\caption{\label{fig:Linear_wkfld}Upper panels: Colormaps of the fields $E_z$ and
  $E_y$ as extracted from the simulation (blue and red), along with
  the laser envelope (orange contour lines). (The laser propagates to the
  right and is polarized along $x$.) The dashed lines indicate the
  position where the curves of the lower panels have been extracted. 
Lower panels: profile of the fields at a given radial
position, as given by \cref{eq:analytical-Ez,eq:analytical-Ey} (dashed
line) or as given by the simulation (red line).}
\end{figure}

Here, the longitudinal laser envelope $a_\ell$ was extracted directly from the
simulation, and the above analytical integrals where carried out
numerically. The resulting predicted fields $E_z$ and $E_y$ are plotted in dashed lines
in the left and right lower panels respectively of
\cref{fig:Linear_wkfld}. 
These predicted curves are compared with
the fields $E_z$ and $E_y$ extracted directly from the simulation (red
lines in the lower panels). These fields from the simulation are also
displayed as colormaps in the upper panels of \cref{fig:Linear_wkfld},
and these colormaps include the line along which the quantities in the
lower panels are plotted (dashed line).

As can be seen on the lower panels of \cref{fig:Linear_wkfld}, the
analytical and simulated curves overlap very precisely, which confirms
the validity of our PIC algorithm. This agreement is all
the more remarkable as the $E_z$ field has been extracted from the cells
that are closest to the axis, a region where quasi-cylindrical algorithms
are typically very noisy.

\section{Advantages over finite-difference algorithms}
\label{sec:advantages}

Since finite-difference algorithms and spectral algorithms are
considerably different, each of them has its own advantages and
shortcomings. 

For instance, spectral algorithms -- including the one described here
-- are generally more difficult to parallelize than finite-difference
codes. Similarly, the implementation of specific boundary
conditions, and of a moving window, is usually more challenging in a
spectral algorithm. On the other hand, due to their high accuracy,
spectral algorithms can avoid a
number of numerical artifacts that are typically present in
finite-difference algorithms. This is quite important as, in a
finite-difference simulation, these artifacts may remain unnoticed
unless additional care is taken in their analysis, and yet they may
affect the physics at stake in an important way.

One example of such an artifact is spurious numerical dispersion, which, as
mentioned in \cite{CowanPRSTAB2013} and in \cref{sec:linear_plasma}, 
can modify the dephasing length of a laser-wakefield accelerator -- a
spurious effect which
may be difficult to discern, especially in the cases where there is no
analytical formula for this length. As shown in
\cref{sec:linear_plasma}, the algorithm described here would not suffer
from this effect, as it is free of spurious numerical dispersion. Similarly, in
this section we describe two other typical situations in which our spectral
quasi-cylindrical algorithm avoids important artifacts, that would otherwise
arise in a finite-difference code. 

\subsection{Suppression of zero-order numerical Cherenkov effect}

A first artifact which is avoided is the zero-order numerical
Cherenkov effect. To zero order, the numerical Cherenkov effect is a 
consequence of the spurious numerical dispersion \cite{GodfreyJCP1974}, and arises 
because some relativistic particles can
travel faster than the numerically-altered velocity of the
electromagnetic waves, in the simulations. This typically causes these
relativistic particles to emit a characteristic spurious
radiation. Recently, it was shown that this spurious radiation
can have a very substantial impact in
lab-frame simulations of laser-wakefield acceleration, 
in particular by spuriously increasing the
emittance of the accelerated beam \cite{LehePRSTAB2013}. It
was also shown that, by modifying the PIC algorithm and its
numerical dispersion relation, the zero-order Cherenkov radiation can be
suppressed. (Notice that there exists also a set of higher-order,
\emph{aliased} numerical Cherenkov effects, which are not as easily
suppressed \cite{GodfreyJCP2013,XuCPC2013,GodfreyJCP2014,
GodfreyIEEE2014,YuJCP2014,YuCPC2015,GodfreyCPC2015}. 
These high-order effects can lead to
disruptive instabilities in boosted-frame simulations, but on the
other hand no such instability was observed in typical lab-frame
simulations of laser-plasma acceleration. 
It is in fact likely that these instabilities may not have enough time to
develop in the case of typical lab-frame simulations.) 

Since the spectral quasi-cylindrical algorithm described here is
dispersion-free, it should not exhibit the zero-order numerical Cherenkov
effect. In order to confirm this prediction, we ran a simulation of
laser-wakefield acceleration with our spectral quasi-cylindrical
algorithm, and compared it with an equivalent simulation that uses 
the finite-difference quasi-cylindrical algorithm of \textsc{Warp}.

In these simulations, a laser pulse with a waist $w_0 = 16\;\mathrm{\mu m}$,
a length $L=10\;\mathrm{\mu m}$, an amplitude $a_0 = 4$ and a
wavelength $\lambda = 0.8\;\mathrm{\mu m}$
is sent into a longitudinally-shaped pre-ionized gas jet. The gas jet
has a 100 $\mathrm{\mu m}$-long rising density gradient at its
entrance, followed by a 200 $\mathrm{\mu m}$ plateau at a density $n_e = 1 \times
 10^{18}\;\mathrm{cm^{-3}}$ and then by a 100 $\mathrm{\mu m}$-long
 downramp, so as to finally reach a density $n_e = 0.5 \times
 10^{18}\;\mathrm{cm^{-3}}$.  Again, the initial temperature of
the plasma is set to 0. The downramp causes the injection of an
 electron beam, which is then accelerated in the subsequent density 
plateau at $n_e = 0.5 \times 10^{18}\;\mathrm{cm^{-3}}$. 
Both the spectral and the finite-difference simulations were run in a moving
window whose longitudinal and transverse dimensions were 160 $\mathrm{\mu
  m}$ and 48 $\mathrm{\mu m}$, and both simulations used a resolution
$\Delta z = 0.032 \; \mathrm{\mu m}$ and $\Delta r = 0.19 \; \mathrm{\mu
  m}$. Again, the finite-difference simulation was run with a timestep
$c\Delta t = 1/\sqrt{1/\Delta z^2 + 2/\Delta r^2}$ while the spectral
simulation was run with $c\Delta t = \Delta z$. Importantly, the
charge and currents are smoothed in both
simulations. The spectral algorithms performs smoothing in spectral
space as described in \cref{sec:current-deposition}, while the finite-difference
algorithm applies a single-pass binomial filter on the spatial grid,
in both the $z$ and $r$ directions.

\begin{figure}[!h]
\centering
\includegraphics[width=\textwidth]{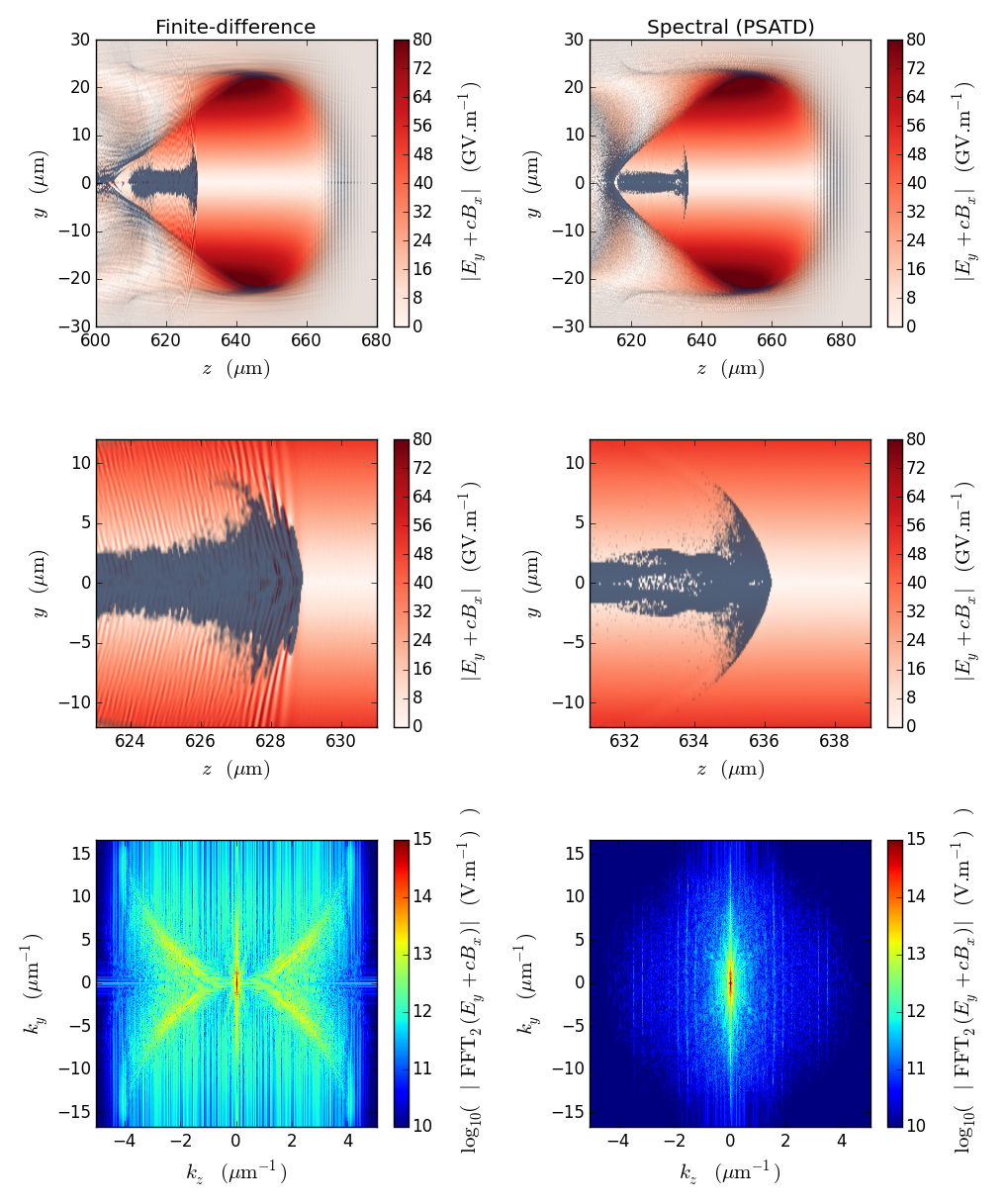}
\caption{\label{fig:Cherenkov}Snapshot of a finite-difference
  simulation (left) and of a spectral simulation (right), at a similar
  physical time. Upper panels: Representation of the bubble in real space; 
the superimposed blue shades represent the electron
  density. Middle panels: More detailed view of the region surrounding the
  bunch. Lower panels: two-dimensional Fourier transform of the
  quantity $E_y + cB_x$ ; the central spot around $k_z=0$ corresponds
  to the physical fields, while the double-parabola in the left panel
  is a signature of the zero-order Cherenkov effect.}
\end{figure}

\Cref{fig:Cherenkov} shows a snapshot of the two
simulations, at a similar physical time. The red colormap represents the quantity
$|E_y+cB_x|$, while the superimposed shades of blue represent the
electron density. (The quantity $E_y+cB_x$ is chosen because it corresponds to the $y$ component of
the Lorentz force $\vec{F} = q\vec{E} + q\vec{v}\times\vec{B}$ felt by
a relativistic electron having $\vec{v} = c\vec{e_z}$.) As can be seen
in the top panels, the global aspect of the bubble and of the injected
bunch is similar in both simulations. In particular, the injected
charge was found to be almost the same (766~pC with the
finite-difference algorithm and 750~pC with the spectral
algorithm). However, a closer look at the bunch in the
finite-difference algorithm (see the middle left panel in
\cref{fig:Cherenkov}) reveals that the bunch emits a
high-frequency radiation. This radiation seems to be due to the
zero-order numerical Cherenkov effect -- an interpretation which is confirmed by
the observation of the corresponding characteristic double-parabola in
$\vec{k}$ space \cite{LehePRSTAB2013,GodfreyJCP2013,XuCPC2013} on the lower left
panel of \cref{fig:Cherenkov}. Importantly, the amplitude of this
unphysical field, in the middle panel of \cref{fig:Cherenkov}, 
happens to be comparable to the focusing
fields inside the bubble, and it can thus potentially affect the bunch
in a substantial manner. On the other hand, the spectral algorithm does not
exhibit this unphysical radiation in real space (see the middle right panel in
\cref{fig:Cherenkov}), nor does it exhibit the corresponding
characteristic pattern in $\vec{k}$ space (see the lower right
panel). This was indeed expected from its dispersion-free property, 
and it confirms the fact that our spectral quasi-cylindrical algorithm is free of
the unphysical artifacts associated with the zero-order numerical
numerical Cherenkov effect.

For completeness, we remark that a similar suppression of the
numerical Cherenkov radiation was 
obtained in \cite{LehePRSTAB2013,CowanPRSTAB2013}, by using 
modified finite-difference algorithms. However, it is important to
note that these algorithms were applicable to a Cartesian PIC code,
but not to a quasi-cylindrical code. An adaptation to
cylindrical geometry is proposed in \cite{LeheThesis}, and was 
shown to efficiently suppress zero-order numerical Cherenkov effect. However, this
adaptation is not, strictly speaking, dispersion-free. 

\subsection{Accurate force of a laser on a copropagating electron}
\label{sec:accurate-laser}

Another case in which our spectral algorithm performs better than
finite-difference algorithms is whenever a relativistic electron bunch
overlaps with a copropagating laser. This situation occurs for
instance in laser-wakefield acceleration, when the accelerated electron bunch
progressively catches up with the laser pulse and may eventually
overlap with it (e.g. \cite{CipicciaNatPhys2011,NemethPRL2008}). It is also the typical
configuration for simulations of free-electron lasers, where the
overlap of the electron bunch and the copropagating laser radiation, inside an
undulator, leads to a growing instability.

Despite the practical importance of this physical configuration, it
was recently shown that standard finite-difference PIC codes tend 
to largely overestimate the force felt by the electrons inside a
copropagating laser (see the appendix of \cite{LehePRSTAB2014}). This
overestimation was shown to be mainly due to the staggering in time 
of the $\vec{E}$ and $\vec{B}$ fields in a standard finite-difference
code. This staggering indeed results in an improper numerical compensation of 
the two terms of the Lorentz force $\vec{F} = q\vec{E} +
q\vec{v}\times\vec{B}$. However, in our spectral quasi-cylindrical
algorithm, the fields $\vec{E}$ and $\vec{B}$ are not staggered in
time, and thus the force on a copropagating electron bunch should be
correct.

\begin{figure}[!h]
\centering
\includegraphics[width=0.5\textwidth]{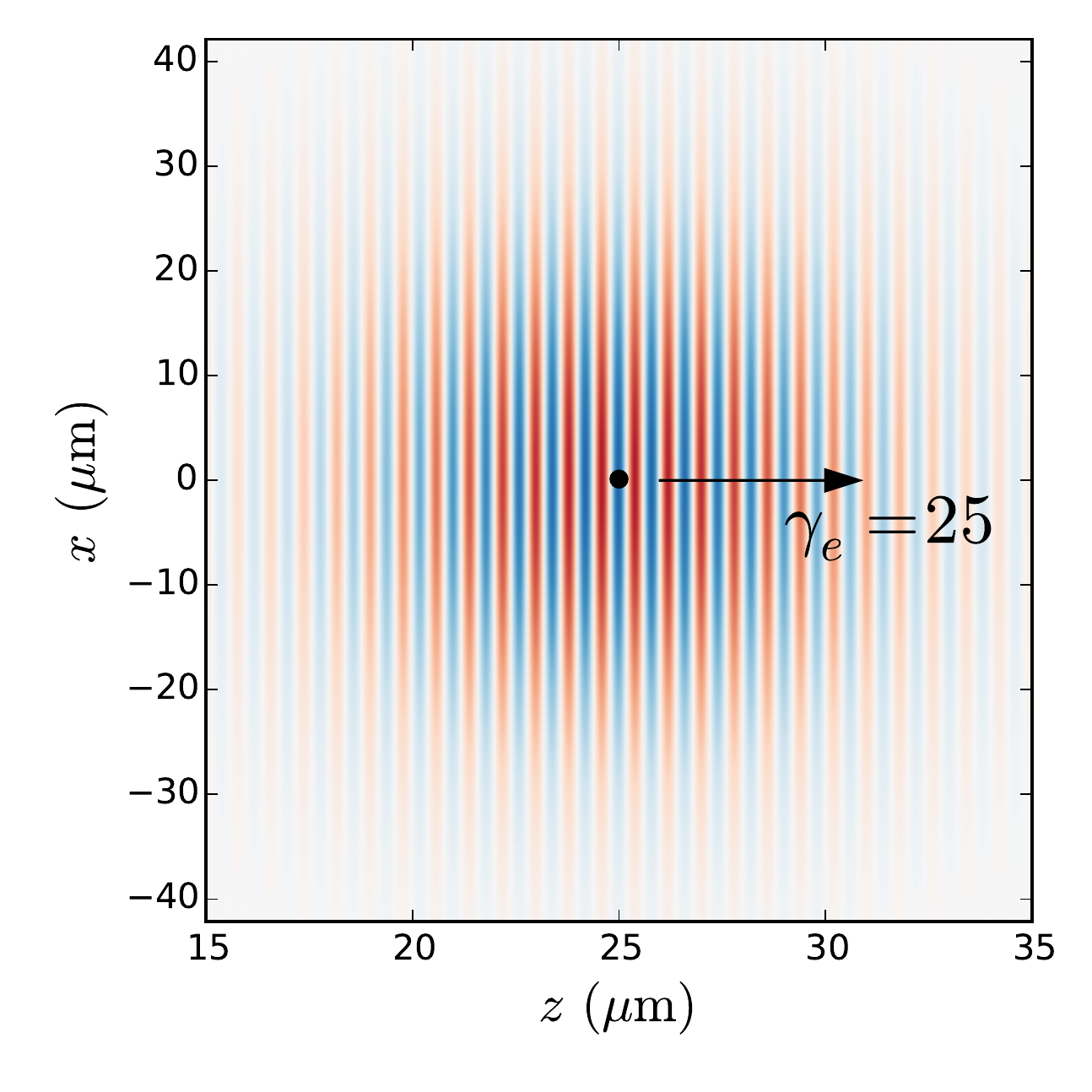}
\caption{\label{fig:Schematic_laser}Representation of the situation
  which is simulated in \cref{sec:accurate-laser}. The red and blue
  colormap corresponds to the electric field of the laser pulse, while
  the black dot corresponds to the electron considered. 
The laser propagates to the right, and is polarized along $x$.}
\end{figure}

To confirm this prediction, we ran a test simulation in which a single
relativistic macroparticle copropagates with a laser pulse. The initial configuration of
the simulation is represented in \cref{fig:Schematic_laser}. The laser
pulse is polarized along $x$ and is characterized by a waist $w_0 = 25\;\mathrm{\mu m}$,
a length $L=7\;\mathrm{\mu m}$, an amplitude $a_0 = 0.2$ and a
wavelength $\lambda = 0.8\;\mathrm{\mu m}$, while the macroparticle
represents a relativistic electron having an initial Lorentz factor
$\gamma_e = 25$. For comparison, the simulation was run, again, with both our spectral
quasi-cylindrical algorithm and the finite-difference
quasi-cylindrical algorithm of \textsc{Warp}. In order to study
numerical convergence, the simulations were run with various
longitudinal resolution $\Delta z$, but with a fixed cell aspect ratio
$\Delta r = 10\Delta z$. The timestep was again $c\Delta t=
1/\sqrt{1/\Delta z^2 + 2/\Delta r^2}$ for the finite-difference
simulation and $c\Delta t = \Delta z$ for the spectral
simulation. Finally, the simulations were run in a moving window with a
longitudinal and transverse size of 40 $\mathrm{\mu m}$ and 60 
$\mathrm{\mu m}$ respectively.

In order to assess the correctness of the force felt by the electron,
we analyze the evolution of its transverse momentum in
time. Analytically, it can be shown from the Lagrangian 
$\mathcal{L} = - \sqrt{1-\vec{\beta}^2} \,mc^2 - e \vec{v} \cdot
\vec{A}$ that the equation of motion along the $x$ axis (i.e. the axis of laser polarization) is 
\begin{equation} \frac{1}{c}\frac{d \,}{dt} \left( \frac{p_x}{m_e c} - a_x \right) =
 - \frac{p_x}{\gamma m_e c}\frac{\partial a_x}{\partial x} \qquad \sim
\frac{a_0^2}{\gamma w_0} \end{equation}
\noindent where $a_x$ is the normalized vector potential of the laser
pulse. The right-hand side corresponds to the ponderomotive force, and,
for the parameters and timescale considered here, it is in fact
negligible. The canonical momentum of the
electron is thus approximately conserved.
\begin{equation} \frac{p_x}{m_e c} - a_x  = const. \end{equation}
\noindent Since, inside the laser pulse, $a_x$ oscillates between the values $-a_0$ and $a_0$,
the quantity $p_x/m_e c$ is predicted to also also oscillate
between $-a_0$ and $a_0$, as the electron progressively
dephases with the laser. 

\begin{figure}[!h]
\centering
\includegraphics[width=0.45\textwidth]{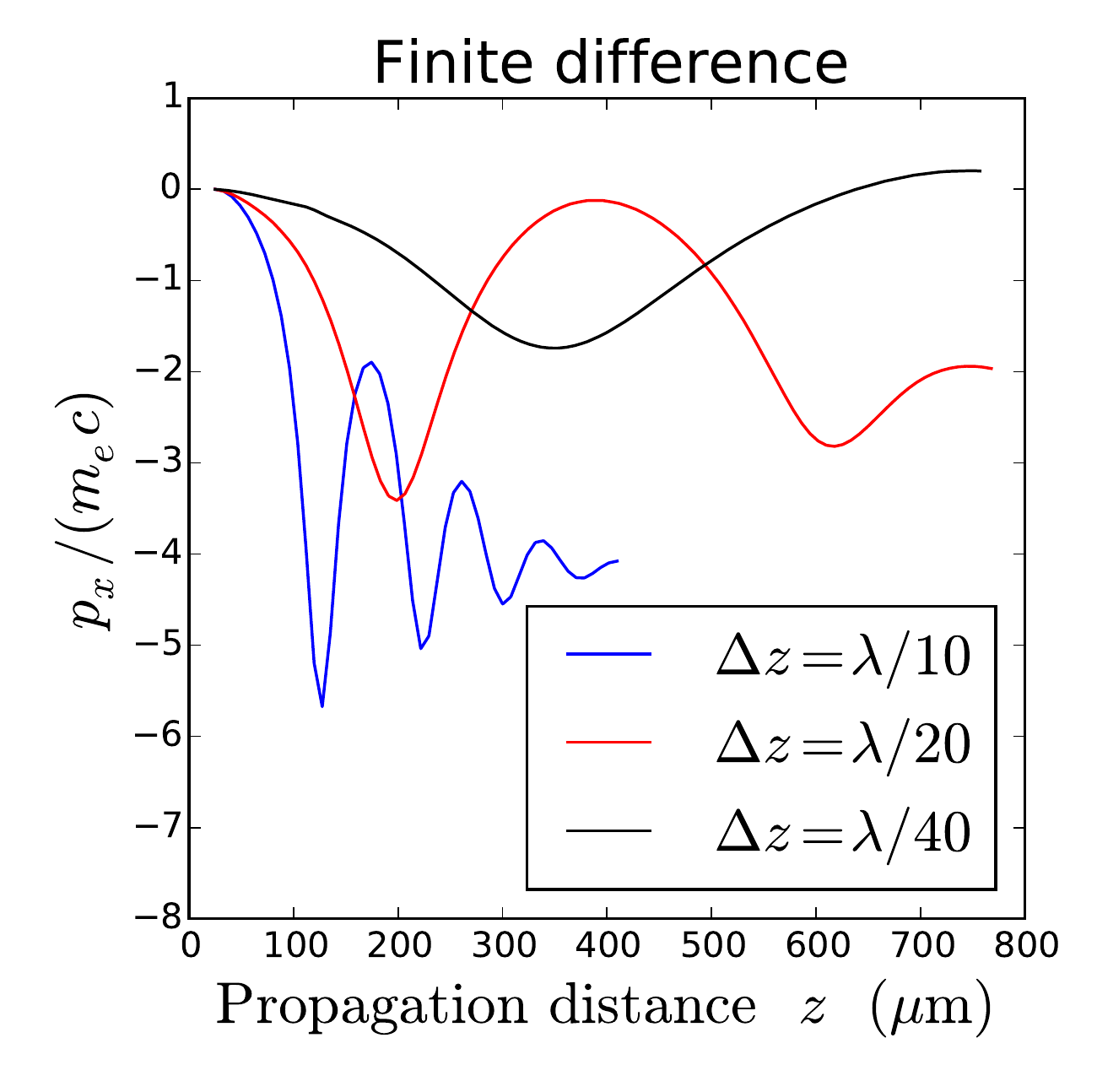}
\includegraphics[width=0.45\textwidth]{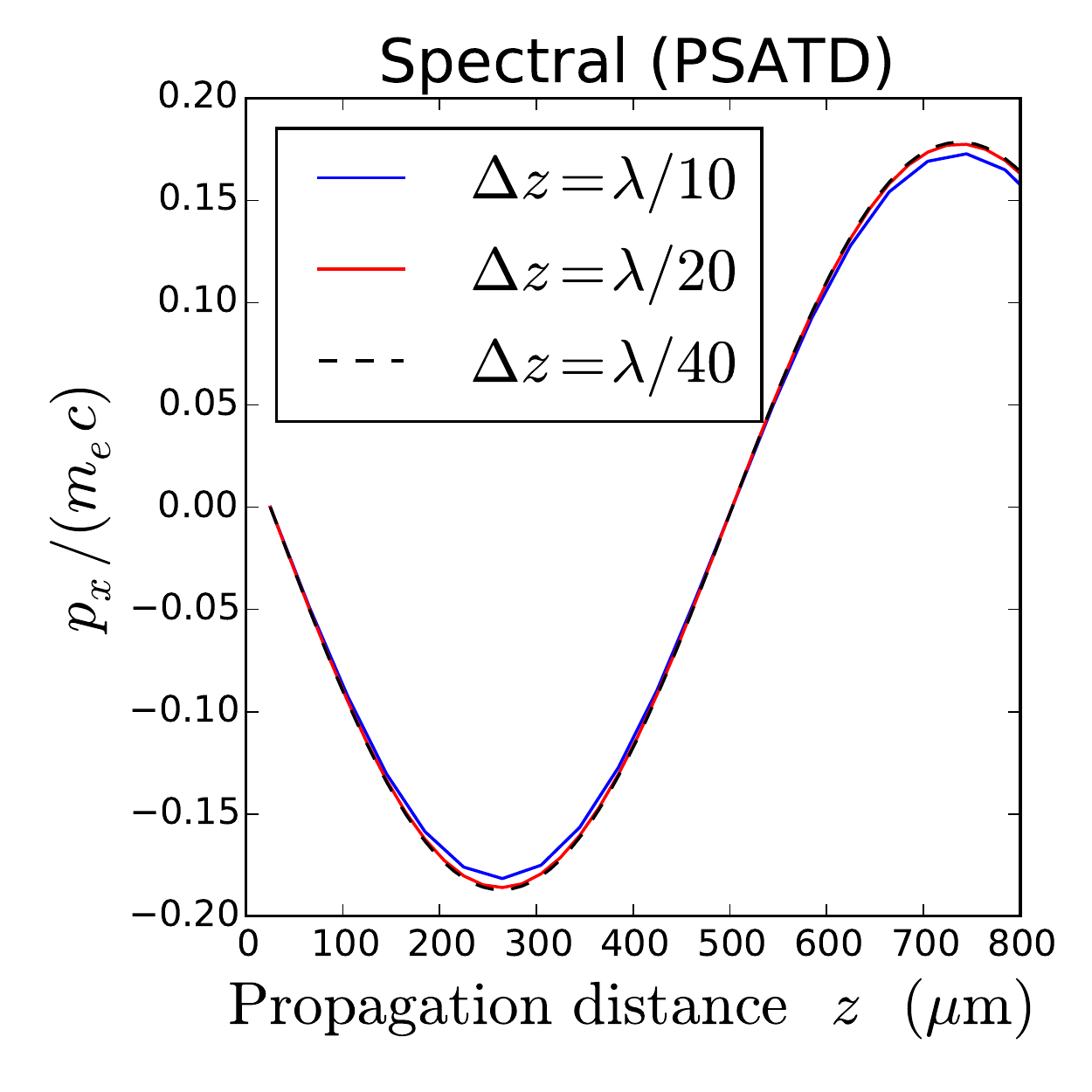}
\caption{\label{fig:Laser}Evolution of the transverse normalized
  momentum of a macroparticle, as it copropagates with a
  laser pulse having $a_0 = 0.2$ (see \cref{fig:Schematic_laser} for a
  schematic representation of the situation simulated). The left and
  right plots correspond to the finite-difference and spectral
  algorithms respectively (note the different vertical scales on the two
  plots). The colored curves correspond to different resolutions for the simulations.}
\end{figure}

Keeping in mind that here initially $a_0=0.2$, one can observe on
\cref{fig:Laser} that the oscillation amplitude of $p_x/m_e c$
is much higher than analytically predicted in the finite-difference
simulation. In addition, these oscillations strongly depend on the
resolution of the simulation, which confirms their spurious nature. These
observations are consistent with those of \cite{LehePRSTAB2014}, and
they are due to the above-mentioned erroneous calculation of the
Lorentz force. On the other hand, in the spectral simulations, the
oscillations of $p_x/m_e c$ exhibits virtually no dependence on the
resolution. In addition, the oscillations of $p_x/m_e c$ have a realistic
amplitude. (The fact that these oscillations do not reach exactly the
value 0.2 may be due to the fact that $a_0$ decreases as the laser
propagates, as a consequence of diffraction.) These observations
confirm that the spectral quasi-cylindrical algorithm properly calculates
the Lorentz force on the electron. This is also consistent with our
expectations, which were based on the fact that $\vec{E}$ and
$\vec{B}$ fields are not staggered in time, in this algorithm.

\section*{Conclusion}

In this article, we derived the equations of a spectral
quasi-cylindrical PIC algorithm, and discussed its numerical
implementation. As explained in the text, because this algorithm is
quasi-cylindrical, it requires much less computational
time and memory than a full 3D algorithm. In addition, the fact
that it is spectral (and uses the PSATD algorithm) prevents a certain number of
artifacts, that are usually present in a finite-difference algorithm.

This second aspect was tested in detail in this article, through
several benchmarks. By comparing simulations with analytical results, 
we showed that our spectral quasi-cylindrical code is free of
spurious numerical dispersion, that it does not exhibit zero-order 
numerical Cherenkov effect in typical lab-frame
simulations, and that it can accurately calculate the force of a
laser on a copropagating electron. By running the
same benchmarks with a finite-difference quasi-cylindrical algorithm, we showed that
these numerical artifacts can, on the contrary, be present in
finite-difference PIC algorithms. 

However, it must not be forgotten that
finite-difference algorithms have advantages of their own, including
easier parallelization and application of boundary conditions. In the
case of our spectral quasi-cylindrical algorithm, parallelization to
several nodes could nonetheless be achieved by following for instance the
method of \cite{VayJCP2013}. This will be the subject of future work. 

\paragraph{Acknowledgements}

We thank Axel Huebl of the PIConGPU team \cite{PIConGPU} for
interesting discussions on the GPU implementation, and Kevin Peters
(U. Hamburg) for contributing to the validation of the code.
We acknowledge computational resources from the Physnet cluster, 
University of Hamburg.

This work was supported by the Director, Office of Science, Office of High Energy Physics, U.S. Dept. of Energy under Contract No. DE-AC02-05CH11231, including from the Laboratory Directed Research and Development (LDRD) funding from Berkeley Lab.

This document was prepared as an account of work sponsored in part by the United States Government. While this document is believed to contain correct information, neither the United States Government nor any agency thereof, nor The Regents of the University of California, nor any of their employees, nor the authors makes any warranty, express or implied, or assumes any legal responsibility for the accuracy, completeness, or usefulness of any information, apparatus, product, or process disclosed, or represents that its use would not infringe privately owned rights. Reference herein to any specific commercial product, process, or service by its trade name, trademark, manufacturer, or otherwise, does not necessarily constitute or imply its endorsement, recommendation, or favoring by the United States Government or any agency thereof, or The Regents of the University of California. The views and opinions of authors expressed herein do not necessarily state or reflect those of the United States Government or any agency thereof or The Regents of the University of California.

\appendix

\section{Derivation of the spectral quasi-cylindrical representation}
\label{sec:CircTrans}

In order to derive the representation
\cref{eq:CircBwTransz,eq:CircBwTransr,eq:CircBwTranst}, we have to
distinguish the Cartesian components (e.g. $E_x$, $E_y$, $E_z$, $B_x$,
$B_y$, $B_z$), which are well-defined everywhere in space and thus have a regular
Fourier representation, from the cylindrical components (e.g. $E_r$,
$E_\theta$, $B_r$, $B_\theta$), which are ill-defined at $r=0$. (For instance, for a
field of the form $\vec{E} = E_0 \vec{e}_x$, the expression of $E_r$
is $E_r = E_0 \cos(\theta)$, which ill-defined at $r=0$, as $\theta$
itself is ill-defined at this position.)

\subsection{Cartesian components}

Let $F_u$ be the Cartesian component of a field $F$ 
(typically $F$ is $E$, $B$ or $J$ and $u$ is $x$, $y$ or $z$). Its Fourier representation
is thus given by \cref{eq:CartBwTrans,eq:CartFwTrans}:
\begin{align}
F_u(\vec{r}) = \frac{1}{(2\pi)^{3}}\Integ{k_x} \,\Integ{k_y}\,
\Integ{k_z} \; \mathcal{F}_u(\vec{k}) e^{i(k_x x + k_y y + k_z z)} \\
\mathcal{F}_u(\vec{k})  = \Integ{x} \,\Integ{y}\,
\Integ{z} \; F_u(\vec{r}) e^{-i(k_x x + k_y y + k_z z)} 
\end{align}
Using the change of variable $k_x=k_\perp\cos(\phi)$, $k_y = k_\perp\sin(\phi)$,
$x=r\cos(\theta)$, $y=r\sin(\theta)$, this becomes
 \begin{align}
F_u(\vec{r}) = \frac{1}{(2\pi)^{3}}\Integ{k_z} \,\RInteg{k_\perp}\,
\TInteg{\phi} \; \mathcal{F}_u(\vec{k})
e^{i(k_\perp r \cos(\theta-\phi) + k_z z)} \\
\mathcal{F}_u(\vec{k})   = \Integ{z} \,\RInteg{r}\,
\Integ{\theta} \; F_u(\vec{r}) e^{-i(k_\perp r \cos(\theta-\phi) + k_z z)} 
\end{align}
Let us now use the relation $e^{ik_\perp r\cos(\theta-\phi)} =
\sum_{m=-\infty}^{\infty} i^m J_m(k_\perp r) e^{im(\phi-\theta)}$
(which is simply another way of writing the well-known relation
$e^{i \alpha \sin \psi} =\sum_{m=-\infty}^{\infty}J_m(\alpha)e^{im\psi}$). The above equations become:
\begin{align}
F_u (\vec{r})  = \sum_{m=-\infty}^{\infty} \frac{1}{(2\pi)^{3}}\Integ{k_z} \,\RInteg{k_\perp }
\TInteg{\phi} \; i^m \mathcal{F}_u(\vec{k}) \;
J_m(k_\perp r) e^{-im(\theta-\phi) + ik_z z} \\
\mathcal{F}_u(\vec{k})   =  \sum_{m=-\infty}^{\infty} \Integ{z} \,\RInteg{r}
\TInteg{\theta} \;\; (-i)^m F_u(\vec{r}) \; J_m(k_\perp r) e^{-im(\phi-\theta) -ik_z z} 
\end{align}
We now define $\spectral{F}_{u,m}(k_z,k_\perp ) = \frac{1}{2\pi}\int_0^{2\pi}
\mathrm{d}\phi \; i^m \mathcal{F}_u(\vec{k})
e^{im\phi}$. This results in the following equations :
\begin{align}
F_u(\vec{r}) =  \frac{1}{(2\pi)^2}\sum_{m=-\infty}^{\infty} \Integ{k_z}
\RInteg{k_\perp }\; \spectral{F}_{u,m}(k_z,k_\perp ) \; J_m(k_\perp r) e^{-im\theta + ik_z z} 
\\
\spectral{F}_{u,m}(k_z,k_\perp ) = \Integ{z} \RInteg{r}
\TInteg{\theta} \;F_u(\vec{r})\; J_m(k_\perp r) e^{-im\theta
 - i k_z z} \label{eq:CircBwTransu}
\end{align}
These equations correspond to \cref{eq:CircBwTransz,eq:CircFwTransz}.

\subsection{Cylindrical components}

Let us now consider fields of the type $E_r$, $B_r$ or $J_r$, which we
denote generally by $F_r$. We have :
\begin{equation} F_r = \cos(\theta) F_x + \sin(\theta) F_y 
= \frac{F_x - iF_y}{2}e^{i\theta} + \frac{F_x +
  iF_y}{2}e^{-i\theta} \end{equation} 
Using \cref{eq:CircBwTransu} leads to
\begin{align} 
F_r =   \frac{1}{(2\pi)^2}\sum_{m=-\infty}^{\infty} \Integ{k_z}\,\RInteg{k_\perp }\;
\left(  J_m(k_\perp r) \frac{\spectral{F}_{x,m} -
    i\spectral{F}_{y,m}}{2}e^{-i(m-1)\theta +ik_z z} + J_m(k_\perp r) \frac{\spectral{F}_{x,m} +   i\spectral{F}_{y,m}}{2}e^{-i(m+1)\theta +
    ik_z z} \right) 
\end{align}
\begin{align}
F_r =  \frac{1}{(2\pi)^2}\sum_{m=-\infty}^{\infty} \Integ{k_z}\,\RInteg{k_\perp }\;
\left(  J_{m+1}(k_\perp r) \frac{\spectral{F}_{x,m+1} -
    i\spectral{F}_{y,m+1}}{2}e^{-im\theta +ik_z z} + J_{m-1}(k_\perp r) \frac{\spectral{F}_{x,m-1} +   i\spectral{F}_{y,m-1}}{2}e^{-im\theta +
    ik_z z} \right) 
\end{align}
where we relabeled the dummy variable $m$ in the above sums. Let us
thus define $\spectral{F}_{-,m} = (\spectral{F}_{x,m-1} +
    i\spectral{F}_{y,m-1})/2$ and $\spectral{F}_{+,m} = (\spectral{F}_{x,m+1} -
    i\spectral{F}_{y,m+1})/2$. This results in:
\begin{equation} 
F_r(\vec{r}) =  \frac{1}{(2\pi)^2}\sum_{m=-\infty}^{\infty} \Integ{k_z}\,\RInteg{k_\perp }\;
\left( \spectral{F}_{+,m}\; J_{m+1}(k_\perp r) +\spectral{F}_{-,m}\; J_{m-1}(k_\perp r)
\right)  e^{-im\theta +ik_z z}
\end{equation}

With the same definitions and the same method, it is also easy to show that:
\begin{equation} 
F_\theta(\vec{r}) = \sum_{m=-\infty}^{\infty} \Integ{k_z}\,\RInteg{k_\perp }\;
i\left( \spectral{F}_{+,m}\; J_{m+1}(k_\perp r) - \spectral{F}_{-,m}\; J_{m-1}(k_\perp r)
\right)  e^{-im\theta +ik_z z}
\end{equation}

\section{Maxwell equations for the spectral coefficients}
\label{sec:SpectMaxwell}

In this section, let us derive the Maxwell equations for the spectral
coefficients \cref{eq:CircMaxwellp,eq:CircMaxwellm,eq:CircMaxwellz}
from the Maxwell equations written in cylindrical coordinates \cref{eq:CircMaxwellr,eq:CircMaxwellt,eq:CircMaxwellzz}.

When replacing the Fourier-Hankel decomposition
(\cref{eq:CircBwTransu,eq:CircBwTransr,eq:CircBwTranst}) in the
Maxwell equations \cref{eq:CircMaxwellr,eq:CircMaxwellt,eq:CircMaxwellzz}, we
first notice that the modes proportional to $e^{-im\theta +ik_z z}$ for different
values of $m$ and $k_z$ are not coupled. These different modes can
thus be treated separately. The same cannot be said of the modes
corresponding to different values of $k_\perp $, since they may be coupled
through the Bessel functions $J_m(k_\perp r)$ and their derivatives. 
In the following, we write only the equations corresponding to $\partial_t \vec{B} =
-\vec{\nabla}\times \vec{E}$, since the equation $c^{-2}\partial_t \vec{E} =
\vec{\nabla}\times\vec{B} - \mu_0 \vec{j}$ can be treated very
similarly. These equations become
\begin{subequations}
\begin{align}
\RInteg{k_\perp } \left[ \; \partial_t \spectral{B}_{+,m}  J_{m+1}(k_\perp r)
  + \partial_t \spectral{B}_{-,m}  J_{m-1}(k_\perp r) \; \right] =&&
\nonumber \\ 
\qquad \RInteg{k_\perp } \left[ \; \spectral{E}_{z,m} \frac{im}{r}
  J_m(k_\perp r) \right.-&\left.
  k_z\spectral{E}_{+,m}J_{m+1}(k_\perp r) + k_z\spectral{E}_{-,m}J_{m-1}(k_\perp r) \;
\right] & \\
\RInteg{k_\perp } \left[ \; \partial_t \spectral{B}_{+,m}  J_{m+1}(k_\perp r)
  - \partial_t \spectral{B}_{-,m}  J_{m-1}(k_\perp r) \; \right] =&&
\nonumber \\
 \RInteg{k_\perp } \left[ \; -k_z\spectral{E}_{+,m}J_{m+1}(k_\perp r)
 \right.-&\left.  k_z\spectral{E}_{-,m}J_{m-1}(k_\perp r) - ik_\perp \spectral{E}_{z,m} J_m'(k_\perp r) \;\right] \\
\RInteg{k_\perp }\; \partial_t \spectral{B}_{z,m}  J_{m}(k_\perp r) =
\RInteg{k_\perp } \left[ \; -ik_\perp
  \spectral{E}_{+,m}\right.&\left(\frac{J_{m+1}(k_\perp r)}{k_\perp r} +
    J_{m+1}'(k_\perp r) \right) + & \nonumber \\
\left. ik_\perp \spectral{E}_{-,m}\left(\frac{J_{m-1}(k_\perp r)}{k_\perp r} +
    J_{m-1}'(k_\perp r) \right) \right.-&\left. \frac{im}{r} \left( E_{+,m} J_{m+1}(k_\perp r) +
    E_{-,m} J_{m-1}(k_\perp r) \right) \;\right] 
\end{align}
\end{subequations}
By taking the sum and difference of the first two equations, and by
rearranging the third equation, we obtain:
\begin{subequations}
\begin{align}
\RInteg{k_\perp } \; 2 \,\partial_t \spectral{B}_{+,m}  J_{m+1}(k_\perp r) =
\RInteg{k_\perp } \left[ \; ik_\perp \spectral{E}_{z,m} \left( \frac{m}{k_\perp r} J_m(k_\perp r) -
    J_m'(k_\perp r) \right) -2 k_z\spectral{E}_{+,m}J_{m+1}(k_\perp r) \;
\right] \\
\RInteg{k_\perp } \; 2\, \partial_t \spectral{B}_{-,m}  J_{m-1}(k_\perp r) \; =
\RInteg{k_\perp } \left[ \;
   ik_\perp \spectral{E}_{z,m} \left( \frac{m}{k_\perp r} J_m(k_\perp r) +
    J_m'(k_\perp r) \right)  + 2k_z\spectral{E}_{-,m}J_{m-1}(k_\perp r) \;
\right] \\
\RInteg{k_\perp }\; \partial_t \spectral{B}_{z,m}  J_{m}(k_\perp r) =
\RInteg{k_\perp } \left[ \; -ik_\perp \spectral{E}_{+,m}\left(\frac{m+1}{k_\perp r}J_{m+1}(k_\perp r) +
    J_{m+1}'(k_\perp r) \right) \right.\nonumber \\
\qquad \left. - ik_\perp \spectral{E}_{-,m}\left(\frac{m-1}{k_\perp r}J_{m-1}(k_\perp r) -
    J_{m-1}'(k_\perp r) \right) \right] 
\end{align}
\end{subequations}
We can now use the relations $\frac{m}{k_\perp r} J_m(k_\perp r) +
    J_m'(k_\perp r) = J_{m-1}(k_\perp r)$ and $\frac{m}{k_\perp r} J_m(k_\perp r) -
    J_m'(k_\perp r) = J_{m+1}(k_\perp r)$ (see relation 9.1.27 in
    \cite{Abramowitz}), and obtain :
\begin{subequations}
\begin{align}
\RInteg{k_\perp } \; 2 \,\partial_t \spectral{B}_{+,m}  J_{m+1}(k_\perp r) =
\RInteg{k_\perp } \left[ \; ik_\perp \spectral{E}_{z,m}\,
    J_{m+1}(k_\perp r) -2 k_z\spectral{E}_{+,m}J_{m+1}(k_\perp r) \;
\right] \\
\RInteg{k_\perp } \; 2\, \partial_t \spectral{B}_{-,m}  J_{m-1}(k_\perp r) \; =
\RInteg{k_\perp } \left[ \;
   ik_\perp \spectral{E}_{z,m} \,
    J_{m-1}(k_\perp r) + 2k_z\spectral{E}_{-,m}J_{m-1}(k_\perp r) \;
\right] \\
\RInteg{k_\perp }\; \partial_t \spectral{B}_{z,m}  J_{m}(k_\perp r) =
\RInteg{k_\perp } \left[ \; -ik_\perp \spectral{E}_{+,m} J_{m}(k_\perp r) - ik_\perp \spectral{E}_{-,m}\,J_{m}(k_\perp r) \right] 
\end{align}
\end{subequations}
Each equation of the above system contains Bessel functions of only one
given order ($m+1$, $m-1$ or $m$). This allows the
different $k_\perp $ components to be separated, since the functions $J_n(k_\perp r)$, for a
fixed $n$ and different values of $k_\perp $, form a basis of the set of real functions:
\begin{subequations}
\begin{align}
2 \,\partial_t \spectral{B}_{+,m} =
ik_\perp \spectral{E}_{z,m} -2 k_z\spectral{E}_{+,m} \\
2\, \partial_t \spectral{B}_{-,m} = ik_\perp \spectral{E}_{z,m} \,
    + 2k_z\spectral{E}_{-,m} \\
 \partial_t \spectral{B}_{z,m} = -ik_\perp \spectral{E}_{+,m}  - ik_\perp \spectral{E}_{-,m}
\end{align}
\end{subequations}

\section{PSATD scheme, in the Fourier-Hankel representation}
\label{sec:PSATDderiv}

We use a scheme very similar to that of \cite{Haber}. In this scheme the currents are considered constant over one timestep, and the charge density is considered linear in time.

\subsection{Expressions for $\spectral{B}_m$}

By combining \cref{eq:CircMaxwellp,eq:CircMaxwellm,eq:CircMaxwellz}
and \cref{eq:SpectCons}, one can find the propagation equations for $B$.
\begin{subequations}
\begin{align}
\partial_t^2 \spectral{B}_{+,m} + c^2(k_\perp ^2+k_z^2) \spectral{B}_{+,m} = 
\mu_0 c^2 \left( - \frac{ik_\perp }{2} \spectral{J}_{z,m} + k_z \spectral{J}_{+,m}
\right) \\
\partial_t^2 \spectral{B}_{-,m} + c^2(k_\perp ^2+k_z^2) \spectral{B}_{-,m} = 
\mu_0 c^2 \left( - \frac{ik_\perp }{2} \spectral{J}_{z,m} - k_z \spectral{J}_{-,m}
\right) \\
\partial_t^2 \spectral{B}_{z,m} + c^2(k_\perp ^2+k_z^2) \spectral{B}_{z,m} =
\mu_0c^2  (ik_\perp  \spectral{J}_{+,m} + ik_\perp \spectral{J}_{-,m} ) 
\end{align}
\end{subequations}
Let us integrate these equations for $t\in [n\Delta t, (n+1)\Delta
t]$. In this interval, $\vec{\spectral{J}}_m(t)$ is constant
and equal to $\vec{\spectral{J}}_m^{n+1/2}$, and thus the right-hand side of the above
equations is constant. Using Green functions, the
general solution of a differential equation of the form 
$\partial_t^2 f + \omega^2 f = g_0$, where $g_0$ is a constant, is 
\begin{equation} f(t) = f(t_0) \cos[\,\omega (t-t_0)\,] + \partial_t f (t_0) \frac{
  \sin[\,\omega (t-t_0)\,]  }{\omega} + \frac{g_0}{\omega^2} (1-
\cos[\,\omega (t-t_0)\,] ) \end{equation}  
We thus use the above expression, with $\omega^2 =c^2(k_\perp^2 +
k_z^2)$, to integrate the fields from $t_0 = n\Delta t$ to $t=(n+1)\Delta t$. In
particular, we use again the Maxwell equations
\cref{eq:CircMaxwellp,eq:CircMaxwellm,eq:CircMaxwellz} to obtain the
expression of $\partial_t \spectral{B}_{m} (t_0)$. This yields:
\begin{subequations}
\begin{align}
\tB{+}{n+1} = \; & C \tB{+}{n} - 
\frac{S}{\omega}\left(-\frac{ik_\perp }{2} \tE{z}{n} + k_z\tE{+}{n}
\right) + \mu_0 c^2\frac{1-C}{\omega^2} \left( -\frac{ik_\perp }{2}
  \tj{z}{n+1/2} + k_z \tj{+}{n+1/2} \right)& \\
\tB{-}{n+1} =\; & C \tB{-}{n} - 
\frac{S}{\omega}\left(- \frac{ik_\perp }{2} \tE{z}{n} - k_z\tE{-}{n}
\right) + \mu_0 c^2\frac{1-C}{\omega^2} \left( - \frac{ik_\perp }{2}
  \tj{z}{n+1/2} - k_z \tj{-}{n+1/2} \right) &\\
\tB{z}{n+1} =\; & C \tB{z}{n} - 
\frac{S}{\omega}\left(ik_\perp \tE{+}{n} + ik_\perp \tE{-}{n}
\right) + \mu_0 c^2\frac{1-C}{\omega^2} \left( ik_\perp
  \tj{+}{n+1/2} + ik_\perp \tj{-}{n+1/2} \right)&
\end{align}
\end{subequations}
where $C = \cos(\omega \Delta t)$ and $S = \sin(\omega \Delta t) $.

\subsection{Expressions for $\spectral{E}_m$}

Similarly, when combining \cref{eq:CircMaxwellp,eq:CircMaxwellm,eq:CircMaxwellz}
and \cref{eq:SpectCons}, the propagation equations for $E$ are:
\begin{subequations}
\begin{align}
\partial_t^2 \spectral{E}_{+,m} + c^2(k_\perp^2 + k_z^2) \spectral{E}_{+,m}
= \frac{c^2}{\epsilon_0} \frac{k_\perp}{2} \spectral{\rho}_m -
\mu_0c^2 \partial_t\spectral{J}_{+,m} \\
\partial_t^2 \spectral{E}_{-,m} + c^2(k_\perp^2 + k_z^2) \spectral{E}_{-,m}
= - \frac{c^2}{\epsilon_0} \frac{k_\perp}{2} \spectral{\rho}_m -
\mu_0c^2 \partial_t\spectral{J}_{-,m} \\
\partial_t^2 \spectral{E}_{z,m} + c^2(k_\perp^2 + k_z^2) \spectral{E}_{z,m}
= - \frac{c^2}{\epsilon_0} i k_z \spectral{\rho}_m -
\mu_0c^2 \partial_t\spectral{J}_{z,m} 
\end{align}
\end{subequations}
Let us again integrate these equations for $t\in [n\Delta t, (n+1)\Delta
t]$. In this interval, $\vec{\spectral{J}}_m(t)$ is constant (thus its time
derivatives drop), and $\spectral{\rho}_m$ is linear in time. As a
consequence the right hand side is proportional to $\trho{n} +
(\trho{n+1}-\trho{n})(t-t_0)/\Delta t$. Using Green functions, the solution of 
$ \partial_t^2 f + \omega^2 f = \trho{n} + (\trho{n+1}-\trho{n})(t-t_0)/\Delta t $ is
\begin{equation} f(t) = f(t_0) \cos[\,\omega(t-t_0)\,] + \partial_tf (t_0)
\frac{\sin[\,\omega(t-t_0)\,]}{\omega} + \trho{n}\frac{1-
  \cos[\,\omega(t-t_0)\,]}{\omega^2} + \frac{\trho{n+1}-\trho{n}}{\omega^2}\left(
  \frac{t-t_0}{\Delta t} - \frac{\sin[\,\omega(t-t_0)\,]}{\omega
    \Delta t}
\right) \end{equation}
which, for $t=t_0 +\Delta t$, reduces to
\begin{equation} f(t_0 +\Delta t) = f(t_0) C + \partial_tf (t_0)
\frac{S}{\omega} 
+ \frac{\trho{n+1}}{\omega^2}\left( 1 - \frac{S}{\omega\Delta t}\right) 
- \frac{\trho{n}}{\omega^2}\left( C - \frac{S}{\omega\Delta t}\right) \end{equation}
Using the above expression, we obtain
\begin{subequations}
\begin{align}
\tE{+}{n+1} = \; & C \tE{+}{n} + 
c^2\frac{S}{\omega}\left(-\frac{ik_\perp }{2} \tB{z}{n} + k_z\tB{+}{n}
- \mu_0 \tj{+}{n+1/2} \right) + \frac{c^2}{\epsilon_0}
\frac{k_\perp}{2}\left[ \frac{\trho{n+1}}{\omega^2}\left(
  1 - \frac{S}{\omega\Delta t}\right) -
\frac{\trho{n}}{\omega^2}\left( C -\frac{S}{\omega\Delta t}\right)\right]  & \\
\tE{-}{n+1} =\; & C \tE{-}{n} +
c^2\frac{S}{\omega}\left(- \frac{ik_\perp }{2} \tB{z}{n} - k_z\tB{-}{n}
- \mu_0 \tj{-}{n+1/2} \right) - \frac{c^2}{\epsilon_0}
\frac{k_\perp}{2}\left[ \frac{\trho{n+1}}{\omega^2}\left(
  1 - \frac{S}{\omega\Delta t}\right) - \frac{\trho{n}}{\omega^2}
\left( C - \frac{S}{\omega\Delta t}\right)\right]  &\\
\tE{z}{n+1} =\; & C \tE{z}{n} + 
c^2\frac{S}{\omega}\left(ik_\perp \tB{+}{n} + ik_\perp \tB{-}{n}
- \mu_0 \tj{z}{n+1/2} \right) - \frac{c^2}{\epsilon_0}
ik_z\left[ \frac{\trho{n+1}}{\omega^2}\left(
  1 - \frac{S}{\omega\Delta t}\right) - \frac{\trho{n}}{\omega^2}
\left( C - \frac{S}{\omega\Delta t}\right)\right]  &
\end{align}
\end{subequations}

\section{Discrete Hankel Transform}
\label{sec:HTMatrix}

\subsection{Calculation of the transformation matrices $M_{n,m}$ and $M'_{n,m}$}

Here, for the Discrete Hankel Transform, we use a transformation
similar to that of \cite{Yu,Guizar,KaiMing}, but
we extend it to the case of an evenly-spaced grid in real space (as opposed to one
that is distributed according to the zeros of the Bessel
function, which would have been inconvenient for current deposition
and field gathering). Moreover, we impose, as much as possible, that the
succession of a Discrete Hankel Transform (DHT) and an Inverse Discrete
Hankel Transform (IDHT) retrieves the initial function. As explained in
the text of the article, we use a matrix formalism for the DHT:
\begin{equation} \mathrm{DHT^m_n}[f] \,(k^m_{\perp,j}) = \sum_{p=0}^{N_r-1} (M_{n,m})_{j,p}
\,f(r_p) \qquad \mathrm{IDHT^m_n}[g] \, (r_j) = \sum_{p=0}^{N_r-1}
(M'_{n,m})_{j,p} \,g(k^m_{\perp,p}) \end{equation}
\noindent where $n$ is the order of the Hankel transform, and where
$m$ is the index of the spectral grid
$k^m_{\perp,j}$ on which the Hankel transform is evaluated. 
In practice, as mentioned in the text, these
transforms are only used in the cases $n=m-1$, $n=m$ or $n=m+1$.

The $N_r\times N_r$ matrices $M_{n,m}$ and $M'_{n,m}$ can be entirely
determined by a set of $N_r^2$ constraints. These
constraints can be found, for instance, by imposing the value of the
DHT for a set of $N_r$ different functions, whose exact analytical Hankel
transforms are known. In our case, we impose that the DHT be equal to the exact analytical
Hankel transform for the eigenmodes of a cavity with
perfectly conducting boundary at $r_{max}$ ($\vec{E}(r_{max},z) =
0$), since these physical eigenmodes should also be eigenmodes of our
PIC cycle. These eigenmodes have the following form:
\begin{subequations}
\begin{align}
E_z \;& \; \propto  J_m(k^m_{\perp,\ell} \,r)\,e^{ik_z z -im\theta} \Theta(r_{max}-r) \\
E_r -i E_\theta \;& \; \propto  J_{m+1}(k^m_{\perp,\ell} \,r) \,e^{ik_z z -im\theta} \Theta(r_{max}-r)\\
E_r +i E_\theta \;& \; \propto  J_{m-1}(k^m_{\perp,\ell} \,r)
                    \,e^{ik_z z -im\theta} \Theta(r_{max}-r) 
\end{align}
\end{subequations}
where $\Theta$ is the Heaviside function and where $k^m_{\perp,\ell} =
\alpha^m_\ell / r_{max}$, with $\alpha^m_\ell$ the $\ell$th positive zero of
the Bessel function of order $m$. 
The exact Hankel transform of these modes, evaluated on the discrete set
$\{k^m_{\perp,j} \}$ reads (see \ref{sec:HT-expression} for a derivation)
\begin{equation} 
\label{eq:HT-expression}
\mathrm{HT}_{n}[ \; J_n(k^m_{\perp,\ell} \,r)  \Theta(r_{max}-r)  \;] \,(k^m_{\perp,j} )
\quad \equiv 2\pi \rInteg \; J_n (k^m_{\perp,j} r) J_n (k^m_{\perp,\ell} r)
\quad = \pi\, r_{max}^2\,[ J_{n+\delta_{n,m}}(\alpha_\ell^m)]^2 \;\delta_{j,\ell} 
\end{equation}
where $n$ is either $m-1$, $m$ or $m+1$. Note that the above relation
is valid for any value of $m$, $j$, $\ell$ and $n$ (provided that $n\in \{m-1,m,m+1\}$), except when $n
\neq 0$, $m \neq 0$ and $\ell = 0$ simultaneously (again, see
\ref{sec:HT-expression} for an explanation). Here we impose that the DHT is consistent with
\cref{eq:HT-expression}, where applicable
\begin{equation}
\label{eq:DHT-matrix}
\mathrm{DHT}^m_{n}[ \; J_n(k^m_{\perp,\ell} \,r) \Theta(r_{max}-r)  \;] \,(k^m_{\perp,j}) 
\quad \equiv \sum_{p=0}^{N_r-1} (M_{n,m})_{j,p}
  J_n(k^m_{\perp,\ell}\,r_p) 
\quad = \pi\, r_{max}^2\,[ J_{n+\delta_{n,m}}(\alpha_\ell^m)]^2 \; \delta_{j,\ell} 
\end{equation}
\noindent and we use the constraints given by \cref{eq:DHT-matrix} to
obtain the matrix $M_{n,m}$.

\paragraph{Case where $n=0$ or $m=0$} In this case,
\cref{eq:DHT-matrix} is applicable for any $j$ and $\ell$ in  $\{ 0,
..., N_r-1 \}$, and thus this provides $N_r^2$ constraints on the matrix
$M_{n,m}$, which allow one to completely determine it. 
In fact, a closer look at \cref{eq:DHT-matrix} shows
that the inverse matrix $M^{-1}_{n,m}$ can be directly extracted from the above relations,
since this inverse matrices is defined by the relation $\sum_p (M_{n,m})_{j,p}
(M^{-1}_{n,m})_{p,\ell} = \delta_{j,\ell}$. Thus, from the above
relations, one can directly infer:
\begin{equation} (M_{n,m}^{-1})_{p,\ell} = \frac{ J_n(k^m_{\perp,\ell}\,r_p) } { \pi\,
  r_{max}^2\,[ J_{n+\delta_{n,m}}(\alpha_\ell^m)]^2  } \end{equation}

\noindent The matrix $M_{n,m}$ can then be extracted, by numerically inverting
the matrices  $M^{-1}_{n,m}$ given by the above expression. In
addition, we impose that $M'_{n,m}$ be exactly equal to
$M^{-1}_{n,m}$, so that the succession of a DHT and IDHT retrieves
exactly the initial function.

\paragraph{Case where $n \neq 0$ and $m \neq 0$} In this case, the
equation \cref{eq:DHT-matrix} is not valid for $\ell=0$, and thus it
provides only $N_r(N_r-1)$ constraints on the matrix $M_{n,m}$, which
is not enough to completely determine it. In this case, we use an
empirical method in which we impose the additional constraints
\begin{equation} (M_{n,m})_{0,p} = 0 \qquad \mathrm{for} \quad p \in \{ 0, ...,
N_r-1 \} \end{equation}
\noindent This constraint is imposed because we choose the amplitude of the
Hankel mode proportional to $J_{n}(k^m_0 r)$ to be 0 in the
simulation. (For $n\neq 0$ and $m \neq 0$, $J_n(k^m_0 r)=0$ for any
$r$, so that the amplitude of the corresponding mode has no physical
meaning whatsoever.) This allows the matrix
$M_{n,m}$ to be entirely determined. In addition, to obtain $M'_{n,m}$, we impose
$ (M'_{n,m})_{p,\ell}=\frac{ J_n(k^m_{\perp,\ell}\,r_p) } { \pi\,
  r_{max}^2\,[ J_{n+\delta_{n,m}}(\alpha_\ell^m)]^2  }$ for any $\ell \in \{1,
..., N_r-1 \}$ (as in the case $n=0$ or $m=0$), but also $(M'_{n,m})_{p,\ell}=0$ for
$\ell=0$. This is done again for consistency with the fact that
$J_n(k^m_0 r)=0$. We note that the
above method gives satisfying results for $m=1$ but not for $m=2$. 
In the future, further work will be done towards a better Hankel transform representation.

\subsection{Derivation of \cref{eq:HT-expression}}
\label{sec:HT-expression}

Let us first remark that, through a change of variable where
$r=r_{max}t$, \cref{eq:HT-expression} is equivalent to
\begin{equation} 
\label{eq:HT-expression2}
\int_0^1 \!\!\! tdt\; J_n (\alpha^m_{\ell} t) J_n (\alpha^m_{j} t)
\quad = \frac{1}{2}\,[
J_{n+\delta_{n,m}}(\alpha_\ell^m)]^2\;\delta_{j,\ell} 
\end{equation}
\noindent and let us prove this equation for different cases,
excluding the case where $n\neq 0$, $m\neq 0$ and $\ell=0$ and in which it is
not valid.

\paragraph{Case where $n=m$ and $\ell \neq 0$} In this case, we use
the relation 11.4.5 of \cite{Abramowitz}. Since $n=m$, and since by
definition $\alpha^m_j$ is the zero of the Bessel function of order $m$, we have
$J_n(\alpha^m_j)= J_n(\alpha^m_\ell)=0$, and the relation 11.4.5 is
thus used in the case $b=0$ and $a=1$ (see relation 11.4.5 in \cite{Abramowitz} for the definition
of the $a$ and $b$ coefficients). This yields
\begin{equation} \int_0^1 \!\!\! tdt\; J_n (\alpha^m_{\ell} t) J_n (\alpha^m_{j} t)
= \frac{1}{2}[ J_n'(\alpha_\ell^m) ]^2 \delta_{j,\ell} = \frac{1}{2}[ J_{n+1}(\alpha_\ell^m) ]^2 \delta_{j,\ell}\end{equation}
\noindent where we further used the relation 9.1.27 in \cite{Abramowitz}, which
reads $ J_n'(\alpha^m_\ell) = -J_{n+1}(\alpha^m_\ell)+
\frac{n}{\alpha^m_\ell} J_{n}(\alpha^m_\ell)$, and took into account
the fact that $J_n (\alpha^m_\ell)=0$ in our case.

\paragraph{Case where $n \neq m$ and $\ell \neq 0$} In this case, $n$
is either $m+1$ or $m-1$, since we restricted ourselves to $n\in \{m-1,m,m+1\}$. Let us
prove the equation \cref{eq:HT-expression2} in the case $n=m+1$ (the
proof for $n=m-1$ being very similar). By
definition, we have $J_m(\alpha^m_\ell) = 0$, and thus from the relation 9.1.27 in
\cite{Abramowitz} $J_{m+1}'(\alpha^m_\ell) + \frac{m+1}{\alpha^m_\ell}
J_{m+1}(\alpha^m_\ell) = 0$. We can thus apply the relation 11.4.5 in
\cite{Abramowitz} with $a=m+1$ and $b=1$, and find 
\begin{equation} \int_0^1 \!\!\! tdt\; J_n (\alpha^m_{\ell} t) J_n (\alpha^m_{j} t) = \frac{1}{2
  (\alpha_\ell^m)^2}\left( (m+1)^2 + (\alpha_\ell^m)^2 - n^2 \right)[
J_{n}(\alpha_\ell^m) ]^2 \delta_{j,\ell} = \frac{1}{2}[J_{n}(\alpha_\ell^m) ]^2 \delta_{j,\ell}  \end{equation}

\paragraph{Case where $m=0$ and $\ell=0$} The proof of the two above
cases used the relation 11.4.5 from \cite{Abramowitz}, which is applicable
as long as $\alpha^m_\ell > 0$. This is the case for $\ell \neq 0$
(i.e. the two above cases) but also for $m=0$ and $\ell=0$
(i.e. $J_0(0) \neq 0$ and thus $\alpha^0_0 > 0$). As a consequence,
the above proofs are also valid for $m=0$ and $\ell=0$.

\paragraph{Case where $n=0$, $m \neq 0$ and $\ell=0$}
In this case, $\alpha^m_\ell = 0$ and thus the relation 11.4.5 from
\cite{Abramowitz} does not apply. However, for $j\neq 0$, the left-hand
side of equation \cref{eq:HT-expression2} reduces to 
\begin{equation} \int_0^1 \!\!\! tdt\; J_n( \alpha_\ell^m t)  J_n( \alpha_j^m t) =
J_0(0) \int_0^1 \!\!\! tdt J_n( \alpha_j^m t) =
\frac{J_0(0)}{(\alpha_j^m)^2} \int_0^{\alpha_\ell^m} \!\!\! tdt
J_n(t)= 0\end{equation}
\noindent where we used relation 11.1.1 in \cite{Abramowitz}. On the
other hand, for $j=0$, one has $\alpha^m_j=\alpha^m_\ell=0$ and thus
\begin{equation} \int_0^1 \!\!\! tdt\; J_n( \alpha_\ell^m t)  J_n( \alpha_j^m t) =
[J_0(0)]^2 \int_0^1 \!\!\! tdt = \frac{1}{2}[J_0(0)]^2 \end{equation}

\bibliography{Bibliography}

\begin{thebibliography}{10}
\expandafter\ifx\csname url\endcsname\relax
  \def\url#1{\texttt{#1}}\fi
\expandafter\ifx\csname urlprefix\endcsname\relax\def\urlprefix{URL }\fi
\expandafter\ifx\csname href\endcsname\relax
  \def\href#1#2{#2} \def\path#1{#1}\fi

\bibitem{Birdsall2004}
C.~Birdsall, A.~Langdon,
  \href{http://books.google.fr/books?id=S2lqgDTm6a4C}{Plasma Physics via
  Computer Simulation}, Series in Plasma Physics, Taylor \& Francis, 2004.
\newline\urlprefix\url{http://books.google.fr/books?id=S2lqgDTm6a4C}

\bibitem{Hockney1988}
R.~Hockney, J.~Eastwood,
  \href{http://books.google.fr/books?id=nTOFkmnCQuIC}{Computer Simulation Using
  Particles}, Taylor \& Francis, 1988.
\newline\urlprefix\url{http://books.google.fr/books?id=nTOFkmnCQuIC}

\bibitem{GodfreyJCP1974}
B.~B. Godfrey, Journal of Computational Physics 15~(4) (1974) 504 -- 521.
\newblock \href {http://dx.doi.org/10.1016/0021-9991(74)90076-X}
  {\path{doi:10.1016/0021-9991(74)90076-X}},
  \href{http://www.sciencedirect.com/science/article/pii/002199917490076X}{[link]}.
\newline\urlprefix\url{http://www.sciencedirect.com/science/article/pii/002199917490076X}

\bibitem{CowanPRSTAB2013}
B.~M. Cowan, D.~L. Bruhwiler, J.~R. Cary, E.~Cormier-Michel, C.~G.~R. Geddes,
  Phys. Rev. ST Accel. Beams 16 (2013) 041303.
\newblock \href {http://dx.doi.org/10.1103/PhysRevSTAB.16.041303}
  {\path{doi:10.1103/PhysRevSTAB.16.041303}},
  \href{http://link.aps.org/doi/10.1103/PhysRevSTAB.16.041303}{[link]}.
\newline\urlprefix\url{http://link.aps.org/doi/10.1103/PhysRevSTAB.16.041303}

\bibitem{godfrey1985iprop}
B.~Godfrey, \href{https://books.google.com/books?id=hos\_OAAACAAJ}{The IPROP
  Three-Dimensional Beam Propagation Code}, Defense Technical Information
  Center, 1985.
\newline\urlprefix\url{https://books.google.com/books?id=hos\_OAAACAAJ}

\bibitem{Lifschitz}
A.~F. Lifschitz, X.~Davoine, E.~Lefebvre, J.~Faure, C.~Rechatin, V.~Malka, J.
  Comput. Phys. 228~(5) (2009) 1803--1814.
\newblock \href {http://dx.doi.org/10.1016/j.jcp.2008.11.017}
  {\path{doi:10.1016/j.jcp.2008.11.017}},
  \href{http://dx.doi.org/10.1016/j.jcp.2008.11.017}{[link]}.
\newline\urlprefix\url{http://dx.doi.org/10.1016/j.jcp.2008.11.017}

\bibitem{Davidson}
A.~{Davidson}, A.~{Tableman}, W.~{An}, F.~S. {Tsung}, W.~{Lu}, J.~{Vieira},
  R.~A. {Fonseca}, L.~O. {Silva}, W.~B. {Mori}, Journal of Computational
  Physics 281 (2015) 1063--1077.
\newblock \href {http://arxiv.org/abs/1403.6890} {\path{arXiv:1403.6890}},
  \href {http://dx.doi.org/10.1016/j.jcp.2014.10.064}
  {\path{doi:10.1016/j.jcp.2014.10.064}}.

\bibitem{EsareyRMP2009}
E.~Esarey, C.~B. Schroeder, W.~P. Leemans, Rev. Mod. Phys. 81 (2009)
  1229--1285.
\newblock \href {http://dx.doi.org/10.1103/RevModPhys.81.1229}
  {\path{doi:10.1103/RevModPhys.81.1229}},
  \href{http://link.aps.org/doi/10.1103/RevModPhys.81.1229}{[link]}.
\newline\urlprefix\url{http://link.aps.org/doi/10.1103/RevModPhys.81.1229}

\bibitem{LinPhysFluids1974}
A.~T. Lin, J.~M. Dawson, H.~Okuda, Physics of Fluids 17~(11) (1974) 1995--2001.
\newblock \href {http://dx.doi.org/http://dx.doi.org/10.1063/1.1694656}
  {\path{doi:http://dx.doi.org/10.1063/1.1694656}},
  \href{http://scitation.aip.org/content/aip/journal/pof1/17/11/10.1063/1.1694656}{[link]}.
\newline\urlprefix\url{http://scitation.aip.org/content/aip/journal/pof1/17/11/10.1063/1.1694656}

\bibitem{Haber}
I.~Haber, R.~Lee, H.~Klein, J.~Boris, 1973.

\bibitem{BunemanJCP1980}
O.~Buneman, C.~Barnes, J.~Green, D.~Nielsen,
  \href{http://www.sciencedirect.com/science/article/pii/0021999180900108}{Principles
  and capabilities of 3-d, e-m particle simulations}, Journal of Computational
  Physics 38~(1) (1980) 1 -- 44.
\newblock \href
  {http://dx.doi.org/http://dx.doi.org/10.1016/0021-9991(80)90010-8}
  {\path{doi:http://dx.doi.org/10.1016/0021-9991(80)90010-8}}.
\newline\urlprefix\url{http://www.sciencedirect.com/science/article/pii/0021999180900108}

\bibitem{DawsonRMP1983}
J.~M. Dawson, Rev. Mod. Phys. 55 (1983) 403--447.
\newblock \href {http://dx.doi.org/10.1103/RevModPhys.55.403}
  {\path{doi:10.1103/RevModPhys.55.403}},
  \href{http://link.aps.org/doi/10.1103/RevModPhys.55.403}{[link]}.
\newline\urlprefix\url{http://link.aps.org/doi/10.1103/RevModPhys.55.403}

\bibitem{Liu}
Q.~H. Liu, Microwave and Optical Technology Letters 15~(3) (1997) 158--165.
\newblock \href
  {http://dx.doi.org/10.1002/(SICI)1098-2760(19970620)15:3<158::AID-MOP11>3.0.CO;2-3}
  {\path{doi:10.1002/(SICI)1098-2760(19970620)15:3<158::AID-MOP11>3.0.CO;2-3}},
  \href{http://dx.doi.org/10.1002/(SICI)1098-2760(19970620)15:3<158::AID-MOP11>3\
  .0.CO;2-3}{[link]}.
\newline\urlprefix\url{http://dx.doi.org/10.1002/(SICI)1098-2760(19970620)15:3<158::AID-MOP11>3\
  .0.CO;2-3}

\bibitem{GodfreyJCP2014}
B.~B. Godfrey, J.-L. Vay, I.~Haber, Journal of Computational Physics 258 (2014)
  689 -- 704.
\newblock \href {http://dx.doi.org/http://dx.doi.org/10.1016/j.jcp.2013.10.053}
  {\path{doi:http://dx.doi.org/10.1016/j.jcp.2013.10.053}},
  \href{http://www.sciencedirect.com/science/article/pii/S0021999113007298}{[link]}.
\newline\urlprefix\url{http://www.sciencedirect.com/science/article/pii/S0021999113007298}

\bibitem{GodfreyIEEE2014}
B.~Godfrey, J.-L. Vay, I.~Haber, Plasma Science, IEEE Transactions on 42~(5)
  (2014) 1339--1344.
\newblock \href {http://dx.doi.org/10.1109/TPS.2014.2310654}
  {\path{doi:10.1109/TPS.2014.2310654}}.

\bibitem{YuJCP2014}
P.~Yu, X.~Xu, V.~K. Decyk, W.~An, J.~Vieira, F.~S. Tsung, R.~A. Fonseca, W.~Lu,
  L.~O. Silva, W.~B. Mori, Journal of Computational Physics 266 (2014) 124 --
  138.
\newblock \href {http://dx.doi.org/http://dx.doi.org/10.1016/j.jcp.2014.02.016}
  {\path{doi:http://dx.doi.org/10.1016/j.jcp.2014.02.016}},
  \href{http://www.sciencedirect.com/science/article/pii/S0021999114001363}{[link]}.
\newline\urlprefix\url{http://www.sciencedirect.com/science/article/pii/S0021999114001363}

\bibitem{YuCPC2015}
P.~Yu, X.~Xu, V.~K. Decyk, F.~Fiuza, J.~Vieira, F.~S. Tsung, R.~A. Fonseca,
  W.~Lu, L.~O. Silva, W.~B. Mori, Computer Physics Communications 192 (2015) 32
  -- 47.
\newblock \href {http://dx.doi.org/http://dx.doi.org/10.1016/j.cpc.2015.02.018}
  {\path{doi:http://dx.doi.org/10.1016/j.cpc.2015.02.018}},
  \href{http://www.sciencedirect.com/science/article/pii/S0010465515000752}{[link]}.
\newline\urlprefix\url{http://www.sciencedirect.com/science/article/pii/S0010465515000752}

\bibitem{VayPRL2007}
J.-L. Vay, Phys. Rev. Lett. 98 (2007) 130405.
\newblock \href {http://dx.doi.org/10.1103/PhysRevLett.98.130405}
  {\path{doi:10.1103/PhysRevLett.98.130405}},
  \href{http://link.aps.org/doi/10.1103/PhysRevLett.98.130405}{[link]}.
\newline\urlprefix\url{http://link.aps.org/doi/10.1103/PhysRevLett.98.130405}

\bibitem{MartinsNatPhys2010}
S.~F. Martins, R.~A. Fonseca, W.~Lu, W.~B. Mori, L.~O. Silva, Nat Phys 6~(4)
  (2010) 311--316.
\newblock \href{http://dx.doi.org/10.1038/nphys1538}{[link]}.
\newline\urlprefix\url{http://dx.doi.org/10.1038/nphys1538}

\bibitem{VayJCP2011}
J.-L. Vay, C.~Geddes, E.~Cormier-Michel, D.~Grote, Journal of Computational
  Physics 230~(15) (2011) 5908 -- 5929.
\newblock \href {http://dx.doi.org/http://dx.doi.org/10.1016/j.jcp.2011.04.003}
  {\path{doi:http://dx.doi.org/10.1016/j.jcp.2011.04.003}},
  \href{http://www.sciencedirect.com/science/article/pii/S0021999111002270}{[link]}.
\newline\urlprefix\url{http://www.sciencedirect.com/science/article/pii/S0021999111002270}

\bibitem{Yuarxiv2015}
P.~{Yu}, X.~{Xu}, A.~{Tableman}, V.~K. {Decyk}, F.~S. {Tsung}, F.~{Fiuza},
  A.~{Davidson}, J.~{Vieira}, R.~A. {Fonseca}, W.~{Lu}, L.~O. {Silva}, W.~B.
  {Mori}, ArXiv e-prints\href {http://arxiv.org/abs/1502.01376}
  {\path{arXiv:1502.01376}}, \href{http://arxiv.org/abs/1502.01376}{[link]}.
\newline\urlprefix\url{http://arxiv.org/abs/1502.01376}

\bibitem{AndriyashJCP2015}
I.~Andriyash, R.~Lehe, V.~Malka, Journal of Computational Physics 282 (2015)
  397 -- 409.
\newblock \href {http://dx.doi.org/http://dx.doi.org/10.1016/j.jcp.2014.11.026}
  {\path{doi:http://dx.doi.org/10.1016/j.jcp.2014.11.026}},
  \href{http://www.sciencedirect.com/science/article/pii/S0021999114007888}{[link]}.
\newline\urlprefix\url{http://www.sciencedirect.com/science/article/pii/S0021999114007888}

\bibitem{Vincenti2015}
H.~{Vincenti}, J.~{Vay}, ArXiv e-prints\href {http://arxiv.org/abs/1507.05572}
  {\path{arXiv:1507.05572}}.

\bibitem{LeeCPC2015}
P.~Lee, J.-L. Vay, Computer Physics Communications 194 (2015) 1 -- 9.
\newblock \href {http://dx.doi.org/http://dx.doi.org/10.1016/j.cpc.2015.04.004}
  {\path{doi:http://dx.doi.org/10.1016/j.cpc.2015.04.004}},
  \href{http://www.sciencedirect.com/science/article/pii/S0010465515001356}{[link]}.
\newline\urlprefix\url{http://www.sciencedirect.com/science/article/pii/S0010465515001356}

\bibitem{Guizar}
M.~Guizar-Sicairos, J.~C. Guti\'{e}rrez-Vega, J. Opt. Soc. Am. A 21~(1) (2004)
  53--58.
\newblock \href {http://dx.doi.org/10.1364/JOSAA.21.000053}
  {\path{doi:10.1364/JOSAA.21.000053}},
  \href{http://josaa.osa.org/abstract.cfm?URI=josaa-21-1-53}{[link]}.
\newline\urlprefix\url{http://josaa.osa.org/abstract.cfm?URI=josaa-21-1-53}

\bibitem{Cree}
M.~Cree, P.~Bones, Computers \& Mathematics with Applications 26~(1) (1993) 1
  -- 12.
\newblock \href
  {http://dx.doi.org/http://dx.doi.org/10.1016/0898-1221(93)90081-6}
  {\path{doi:http://dx.doi.org/10.1016/0898-1221(93)90081-6}},
  \href{http://www.sciencedirect.com/science/article/pii/0898122193900816}{[link]}.
\newline\urlprefix\url{http://www.sciencedirect.com/science/article/pii/0898122193900816}

\bibitem{Yu}
L.~Yu, M.~Huang, M.~Chen, W.~Chen, W.~Huang, Z.~Zhu, Opt. Lett. 23~(6) (1998)
  409--411.
\newblock \href {http://dx.doi.org/10.1364/OL.23.000409}
  {\path{doi:10.1364/OL.23.000409}},
  \href{http://ol.osa.org/abstract.cfm?URI=ol-23-6-409}{[link]}.
\newline\urlprefix\url{http://ol.osa.org/abstract.cfm?URI=ol-23-6-409}

\bibitem{Siegman}
A.~E. Siegman, Opt. Lett. 1~(1) (1977) 13--15.
\newblock \href {http://dx.doi.org/10.1364/OL.1.000013}
  {\path{doi:10.1364/OL.1.000013}},
  \href{http://ol.osa.org/abstract.cfm?URI=ol-1-1-13}{[link]}.
\newline\urlprefix\url{http://ol.osa.org/abstract.cfm?URI=ol-1-1-13}

\bibitem{KaiMing}
Y.~Kai-Ming, W.~Shuang-Chun, C.~Lie-Zun, W.~You-Wen, H.~Yong-Hua, Chinese
  Physics B 18~(9) (2009) 3893.
\newblock \href{http://stacks.iop.org/1674-1056/18/i=9/a=046}{[link]}.
\newline\urlprefix\url{http://stacks.iop.org/1674-1056/18/i=9/a=046}

\bibitem{VayPoP2008}
J.-L. Vay, Physics of Plasmas (1994-present) 15~(5) (2008) 056701.
\newblock \href {http://dx.doi.org/http://dx.doi.org/10.1063/1.2837054}
  {\path{doi:http://dx.doi.org/10.1063/1.2837054}},
  \href{http://scitation.aip.org/content/aip/journal/pop/15/5/10.1063/1.2837054}{[link]}.
\newline\urlprefix\url{http://scitation.aip.org/content/aip/journal/pop/15/5/10.1063/1.2837054}

\bibitem{Esirkepov}
T.~Esirkepov, Computer Physics Communications 135~(2) (2001) 144 -- 153.
\newblock \href
  {http://dx.doi.org/http://dx.doi.org/10.1016/S0010-4655(00)00228-9}
  {\path{doi:http://dx.doi.org/10.1016/S0010-4655(00)00228-9}},
  \href{http://www.sciencedirect.com/science/article/pii/S0010465500002289}{[link]}.
\newline\urlprefix\url{http://www.sciencedirect.com/science/article/pii/S0010465500002289}

\bibitem{VayJCP2013}
J.-L. Vay, I.~Haber, B.~B. Godfrey, Journal of Computational Physics 243 (2013)
  260 -- 268.
\newblock \href {http://dx.doi.org/http://dx.doi.org/10.1016/j.jcp.2013.03.010}
  {\path{doi:http://dx.doi.org/10.1016/j.jcp.2013.03.010}},
  \href{http://www.sciencedirect.com/science/article/pii/S0021999113001873}{[link]}.
\newline\urlprefix\url{http://www.sciencedirect.com/science/article/pii/S0021999113001873}

\bibitem{FFTW}
\href{http://www.fftw.org/}{Fftw}.
\newline\urlprefix\url{http://www.fftw.org/}

\bibitem{BLAS}
\href{http://www.netlib.org/blas/}{Blas}.
\newline\urlprefix\url{http://www.netlib.org/blas/}

\bibitem{Numba}
\href{http://numba.pydata.org/}{Numba}.
\newline\urlprefix\url{http://numba.pydata.org/}

\bibitem{Warpref}
J.-L. Vay, D.~P. Grote, R.~H. Cohen, A.~Friedman, Computational Science \&
  Discovery 5~(1) (2012) 014019.
\newblock \href{http://stacks.iop.org/1749-4699/5/i=1/a=014019}{[link]}.
\newline\urlprefix\url{http://stacks.iop.org/1749-4699/5/i=1/a=014019}

\bibitem{YuIPAC2015}
P.~{Yu}, A.~{Davidson}, A.~{Tableman}, V.~K. {Decyk}, F.~S. {Tsung}, W.~B.
  {Mori}, 2015.

\bibitem{Esarey1999}
E.~Esarey, W.~P. Leemans, Phys. Rev. E 59 (1999) 1082--1095.
\newblock \href {http://dx.doi.org/10.1103/PhysRevE.59.1082}
  {\path{doi:10.1103/PhysRevE.59.1082}},
  \href{http://link.aps.org/doi/10.1103/PhysRevE.59.1082}{[link]}.
\newline\urlprefix\url{http://link.aps.org/doi/10.1103/PhysRevE.59.1082}

\bibitem{Karkkainen}
M.~Karkkainen, E.~Gjonaj, T.~Lau, T.~Weiland, Vol.~1, Chamonix, France, 2006.

\bibitem{Pukhov}
A.~Pukhov, Journal of Plasma Physics 61~(03) (1999) 425--433.
\newblock \href {http://dx.doi.org/null} {\path{doi:null}}.

\bibitem{Nuter}
R.~Nuter, M.~Grech, P.~Gonzalez~de Alaiza~Martinez, G.~Bonnaud,
  E.~d'Humi\`eres, The European Physical Journal D 68~(6).
\newblock \href {http://dx.doi.org/10.1140/epjd/e2014-50162-y}
  {\path{doi:10.1140/epjd/e2014-50162-y}},
  \href{http://dx.doi.org/10.1140/epjd/e2014-50162-y}{[link]}.
\newline\urlprefix\url{http://dx.doi.org/10.1140/epjd/e2014-50162-y}

\bibitem{LehePRSTAB2013}
R.~Lehe, A.~Lifschitz, C.~Thaury, V.~Malka, X.~Davoine, Phys. Rev. ST Accel.
  Beams 16 (2013) 021301.
\newblock \href {http://dx.doi.org/10.1103/PhysRevSTAB.16.021301}
  {\path{doi:10.1103/PhysRevSTAB.16.021301}},
  \href{http://link.aps.org/doi/10.1103/PhysRevSTAB.16.021301}{[link]}.
\newline\urlprefix\url{http://link.aps.org/doi/10.1103/PhysRevSTAB.16.021301}

\bibitem{GodfreyJCP2013}
B.~B. Godfrey, J.-L. Vay, Journal of Computational Physics 248 (2013) 33 -- 46.
\newblock \href {http://dx.doi.org/http://dx.doi.org/10.1016/j.jcp.2013.04.006}
  {\path{doi:http://dx.doi.org/10.1016/j.jcp.2013.04.006}},
  \href{http://www.sciencedirect.com/science/article/pii/S0021999113002556}{[link]}.
\newline\urlprefix\url{http://www.sciencedirect.com/science/article/pii/S0021999113002556}

\bibitem{XuCPC2013}
X.~Xu, P.~Yu, S.~F. Martins, F.~S. Tsung, V.~K. Decyk, J.~Vieira, R.~A.
  Fonseca, W.~Lu, L.~O. Silva, W.~B. Mori, Computer Physics Communications
  184~(11) (2013) 2503 -- 2514.
\newblock \href {http://dx.doi.org/http://dx.doi.org/10.1016/j.cpc.2013.07.003}
  {\path{doi:http://dx.doi.org/10.1016/j.cpc.2013.07.003}},
  \href{http://www.sciencedirect.com/science/article/pii/S0010465513002312}{[link]}.
\newline\urlprefix\url{http://www.sciencedirect.com/science/article/pii/S0010465513002312}

\bibitem{GodfreyCPC2015}
B.~B. Godfrey, J.-L. Vay, Computer Physics Communications (2015) --\href
  {http://dx.doi.org/http://dx.doi.org/10.1016/j.cpc.2015.06.008}
  {\path{doi:http://dx.doi.org/10.1016/j.cpc.2015.06.008}},
  \href{http://www.sciencedirect.com/science/article/pii/S0010465515002532}{[link]}.
\newline\urlprefix\url{http://www.sciencedirect.com/science/article/pii/S0010465515002532}

\bibitem{LeheThesis}
R.~Lehe, Theses, {Ecole Polytechnique} (Jul. 2014).
\newblock \href{https://pastel.archives-ouvertes.fr/tel-01088398}{[link]}.
\newline\urlprefix\url{https://pastel.archives-ouvertes.fr/tel-01088398}

\bibitem{CipicciaNatPhys2011}
S.~Cipiccia, M.~R. Islam, B.~Ersfeld, R.~P. Shanks, E.~Brunetti, G.~Vieux,
  X.~Yang, R.~C. Issac, S.~M. Wiggins, G.~H. Welsh, M.-P. Anania, D.~Maneuski,
  R.~Montgomery, G.~Smith, M.~Hoek, D.~J. Hamilton, N.~R.~C. Lemos, D.~Symes,
  P.~P. Rajeev, V.~O. Shea, J.~M. Dias, D.~A. Jaroszynski, Nat Phys 7~(11)
  (2011) 867--871.
\newblock \href{http://dx.doi.org/10.1038/nphys2090}{[link]}.
\newline\urlprefix\url{http://dx.doi.org/10.1038/nphys2090}

\bibitem{NemethPRL2008}
K.~N\'emeth, B.~Shen, Y.~Li, H.~Shang, R.~Crowell, K.~C. Harkay, J.~R. Cary,
  Phys. Rev. Lett. 100 (2008) 095002.
\newblock \href {http://dx.doi.org/10.1103/PhysRevLett.100.095002}
  {\path{doi:10.1103/PhysRevLett.100.095002}},
  \href{http://link.aps.org/doi/10.1103/PhysRevLett.100.095002}{[link]}.
\newline\urlprefix\url{http://link.aps.org/doi/10.1103/PhysRevLett.100.095002}

\bibitem{LehePRSTAB2014}
R.~Lehe, C.~Thaury, E.~Guillaume, A.~Lifschitz, V.~Malka, Phys. Rev. ST Accel.
  Beams 17 (2014) 121301.
\newblock \href {http://dx.doi.org/10.1103/PhysRevSTAB.17.121301}
  {\path{doi:10.1103/PhysRevSTAB.17.121301}},
  \href{http://link.aps.org/doi/10.1103/PhysRevSTAB.17.121301}{[link]}.
\newline\urlprefix\url{http://link.aps.org/doi/10.1103/PhysRevSTAB.17.121301}

\bibitem{PIConGPU}
M.~Bussmann, H.~Burau, T.~E. Cowan, A.~Debus, A.~Huebl, G.~Juckeland, T.~Kluge,
  W.~E. Nagel, R.~Pausch, F.~Schmitt, U.~Schramm, J.~Schuchart, R.~Widera, in:
  Proceedings of the International Conference on High Performance Computing,
  Networking, Storage and Analysis, SC '13, ACM, New York, NY, USA, 2013, pp.
  5:1--5:12.
\newblock \href {http://dx.doi.org/10.1145/2503210.2504564}
  {\path{doi:10.1145/2503210.2504564}},
  \href{http://doi.acm.org/10.1145/2503210.2504564}{[link]}.
\newline\urlprefix\url{http://doi.acm.org/10.1145/2503210.2504564}

\bibitem{Abramowitz}
M.~Abramowitz, I.~Stegun, Handbook of Mathematical Functions, Dover
  Publications, 1972.

\end{thebibliography}
\bibliographystyle{elsarticle-num}

\end{document}